\documentclass[%
reprint,
showpacs,reprintnumbers,
 amsmath,amssymb,
 aps,
prl,
superscriptaddress,
showpacs, showkeys]{revtex4-1}
\usepackage{amsmath}
 \usepackage{subfigure}
\usepackage{textgreek}
\usepackage{subfloat}
\usepackage{color}
\usepackage{breqn}
\usepackage{graphicx}% Include figure files
\usepackage{dcolumn}% Align table columns on decimal point
\usepackage{bm}% bold math
\usepackage{hyperref}% add hypertext capabilities
\usepackage{textcomp}% ufor single quotes
\usepackage{xr}
\hypersetup{bookmarksnumbered, pdfpagemode=UseOutlines, 
colorlinks=true, citecolor=blue, filecolor=blue, linkcolor=blue, urlcolor=blue}

 \newcommand{\beginsupplement}{%
         \setcounter{table}{0}
         \renewcommand{\thetable}{S\arabic{table}}%
         \setcounter{figure}{0}
         \renewcommand{\thefigure}{S\arabic{figure}}
         \setcounter{equation}{0}
         \renewcommand{\theequation}{S\arabic{equation}}%   
      }

 \makeatletter
 \let\cat@comma@active\@empty
 \makeatother
\graphicspath{/home/sid/Dropbox/6th_paper_WS2_susbstrate/paper_fig
}

\begin{document}

\preprint{}

% \title{Efficient spin-valley coupling in spin-orbit coupled bilayer-graphene}
\title{Large spin-relaxation anisotropy in bilayer-graphene/WS$_{\text{2}}$ heterostructures}

\author{S. Omar} 
\thanks{corresponding author}
\email{s.omar@rug.nl}
\affiliation{The Zernike Institute for Advanced Materials University of Groningen Nijenborgh 4 9747 AG, Groningen, The Netherlands}
\author{B.N. Madhushankar}
\affiliation{The Zernike Institute for Advanced Materials University of Groningen Nijenborgh 4 9747 AG, Groningen, The Netherlands}%
\author{B.J. van Wees}
\affiliation{The Zernike Institute for Advanced Materials University of Groningen Nijenborgh 4 9747 AG, Groningen, The Netherlands}%
\date{\today}% It is always \today, today,

\begin{abstract}
We study spin-transport in bilayer-graphene (BLG), spin-orbit coupled to a tungsten di sulfide (WS$_{\text{2}}$) substrate, and measure a record spin lifetime anisotropy $\sim$ 40-70, i.e. ratio between the out-of-plane $\tau_{\perp}$ and in-plane spin relaxation time $\tau_{||}$. We control the injection and detection of in-plane and out-of-plane spins via the shape-anisotropy of the ferromagnetic electrodes.  We estimate $\tau_{\perp}\sim$ 1-2 ns via Hanle measurements at high perpendicular magnetic fields and via a new tool we develop: Oblique Spin Valve measurements. Using Hanle spin-precession experiments we find a low $\tau_{||} \sim$ 30 ps in the electron-doped regime which only weakly depends on the carrier density in the BLG and conductivity of the underlying WS$_{\text{2}}$, indicating proximity-induced spin-orbit coupling (SOC) in the BLG. Such high $\tau_{\perp}$ and spin lifetime anisotropy are clear signatures of strong spin-valley coupling for out-of-plane spins in BLG/WS$_{\text{2}}$ systems in the presence of SOC, and unlock the potential of BLG/transition metal dichalcogenide heterostructures for developing future spintronic applications.
% \begin{description}
% %\item[Usage]
% %Secondary publications and information retrieval purposes.
% \item[PACS numbers]
%  \verb+85.75.-d+, \verb+73.22.Pr+, \verb+75.76.j+ 
% %\item[Structure]
% %You may use the \texttt{description} environment to structure your abstract;
% %use the optional argument of the \verb+\item+ command to give the category of each item.
% \end{description}
\end{abstract}

%Use showkeys class option if keyword
\maketitle

% \section{introduction}

% Graphene (Gr), a representative of two-dimensional (2D) material family \cite{geim_van_2013}, is an ideal material for spintronic applications due to its  excellent spin transport properties \cite{ingla-aynes_$24ensuremath-ensuremathmumathrmm$_2015,drogeler_spin_2016}. In addition, graphene's charge and spin transport properties can be engineered when it is glued to other 2D-materials such as insulating boron-nitride \cite{dean_boron_2010,gurram_bias_2017} and semiconducting transition metal dichalcogenides (TMD) \cite{omar_spin_2018, wang_origin_2016} via weak van der Waals interaction between the layers. 
Graphene (Gr) in contact with a transition metal dichalcogenide (TMD), having high intrinsic spin-orbit coupling (SOC) offers a unique platform where the charge transport properties in Gr are well preserved due to the weak van der Waals interaction between the two materials. However, the spin transport properties are greatly affected due to the TMD-proximity induced SOC in graphene \cite{wang_origin_2016,gmitra_graphene_2015,omar_spin_2018}.  
% In addition, graphene's charge and spin transport properties can be engineered when it is glued to other 2D-materials such as insulating boron-nitride \cite{dean_boron_2010,gurram_bias_2017} and semiconducting transition metal dichalcogenides (TMD) \cite{omar_spin_2018, wang_origin_2016} via weak van der Waals interaction between the layers. 
At the Gr/TMD interface, the spatial inversion symmetry is broken, and the graphene sublattices having K(K') valleys experience different crystal potentials and spin-orbit coupling magnitudes from the underlying TMD. The electron-spin degree of freedom and its interaction with other properties such as valley pseudospins in the presence of SOC provide access to spintronic phenomena such as spin-valley coupling \cite{xiao_valley-contrasting_2007,leutenantsmeyer_observation_2018, xu_strong_2018,zihlmann_large_2018, cummings_giant_2017, ghiasi_large_2017}, spin-Hall effect \cite{safeer_room-temperature_2019,garcia_spin_2017}, (inverse)  Rashba-Edelstein effect \cite{song_observation_2017, soumyanarayanan_emergent_2016, isasa_origin_2016, sanchez_spin--charge_2013,shen_microscopic_2014} and even topologically protected spin-states \cite{kane_$z_2$_2005, yang_tunable_2016,frank_protected_2018,du_robust_2018, island_spinorbit-driven_2019} which are not possible to realize in pristine graphene. The mentioned effects are sought after for realizing enhancement and electric field control of SOC \cite{khoo_-demand_2017,gmitra_proximity_2017,afzal_gate_2018,ye_electric_2017,wang_origin_2016,omar_spin_2018,omar_graphene-$mathrmws_2$_2017}, efficient charge-current to spin-current conversion and vice versa \cite{safeer_room-temperature_2019,offidani_optimal_2017, huang_anomalous_2017,ando_spin_2016}, which will be the building blocks  for developing novel spintronic applications \cite{soumyanarayanan_emergent_2016,gurram_electrical_2018}. 

Experiments on Gr/TMD systems confirm the presence of enhanced spin-orbit coupling \cite{omar_spin_2018, wang_strong_2015} and the anisotropy in the in-plane ($\tau_{||}$) and out-of-plane ($\tau_{\perp}$) spin relaxation times  \cite{ghiasi_large_2017,benitez_strongly_2018,zihlmann_large_2018} in single layer graphene. Recent theoretical studies \cite{khoo_-demand_2017,gmitra_proximity_2017} predict that due to the special band-structure of bilayer-graphene on a TMD substrate, it is expected to show a larger spin-relaxation anisotropy $\eta=\frac{\tau_{\perp}}{\tau_{||}}$ even up to 10000 \cite{gmitra_proximity_2017}, which is approximately 1000 times higher than the highest reported $\eta$ values for single-layer graphene-TMD heterostructures \cite{ghiasi_large_2017, zhu_modeling_2018}. As explained in Ref.~\cite{gmitra_proximity_2017}, a finite band-gap opens up in bilayer-graphene (BLG) in presence of a built-in electric field at the BLG/TMD interface, which can be tuned via an external electric field. The BLG valence (conduction) band is formed via the carbon atom orbitals at the bottom (top) layer. As a consequence, due to the closer proximity of the bottom BLG layer with the TMD, the BLG valence band has almost two order higher magnitude of SOC of spin-valley coupling character than the SOC in the conduction band. This modulation in the SOC can be accessed in two ways: either by the application of a back-gate voltage by tuning the Fermi energy or via the electric-field by changing the sign of the orbital-gap. Depending on whether the graphene is hole or electron doped, and the magnitude of the electric field at the interface, BLG can therefore exhibit the effect of spin-valley coupling in the magnitude of spin-relaxation anisotropy ratio $\eta$.   

In this letter, we report the transport of both in-plane and out-of-plane spins in BLG supported on a TMD substrate, i.e. tungsten disulfide (WS$_2$). We inject and detect the out-of-plane spins in graphene via a purely electrical method by exploiting the magnetic shape anisotropy of the ferromagnetic electrodes at high magnetic fields \cite{tombros_anisotropic_2008, popinciuc_electronic_2009, guimaraes_controlling_2014}, in contrast with the optical injection of out-of-plane spins into Gr/TMD systems in refs. \cite{avsar_optospintronics_2017,luo_opto-valleytronic_2017}. We extract $\tau_{\perp} \sim$ 1 ns-2 ns, which results in $\eta=\frac{\tau_{\perp}}{\tau_{||}}\sim$40-70 via two independent methods; Hanle measurements at high perpendicular magnetic field and a newly developed tool \textit{Oblique Spin Valve} measurements. Such large $\eta$ confirms the existence of strong spin-valley coupling for the out-of-plane spins in BLG/TMD systems. We find a weak modulation in both $\tau_{||}$ and $\tau_{\perp}$  as a function of charge carrier density in the electron-doped regime in the BLG. $\tau_{||}$ varies from 15-30 ps, with such short values indicating the presence of a very strong spin-orbit coupling in the BLG, induced by the WS$_2$ substrate.

Bilayer-graphene/WS$_2$ samples are prepared on a SiO$_2$/Si substrate (thickness $t_{\text{SiO}_2}\sim$500 nm) via a dry pick-up transfer method  \cite{zomer_fast_2014} (see Supplemental Material for fabrication details). 
\begin{figure}
%  \begin{subfigure}{0.45\linewidth}
 \includegraphics[scale=1]{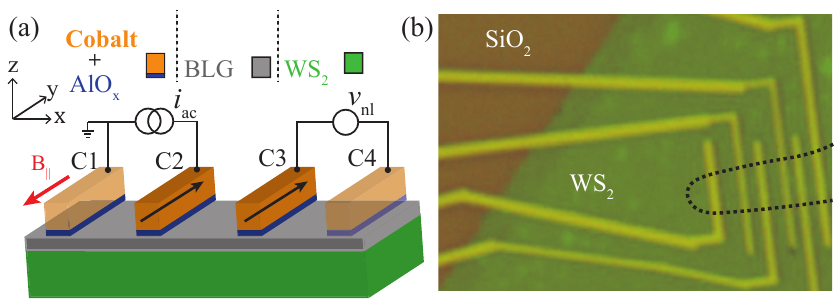}
 \caption{\label{fig1} (a) Nonlocal spin-transport measurement scheme. The ferromagnetic electrodes C2-C3 are premagnetized along the y-axis by applying an in-plane magnetic field. The outer electrodes C1 and C4 act as reference electrodes.  (b) An optical image of a part of  WS$_{\text{2}}$/BLG sample (stack A) where the measurements are performed. The BLG is outlined with a black dashed line which extends further to the right.} 
\end{figure}
We study two bottom-WS$_{\text{2}}$/BLG samples (thickness $t_{\text{WS$_2$}}\sim$ 3 nm), labeled as stack A and stack B, and present the data from the left region of stack A (Fig.~\ref{fig1}(b)) as a representative sample. Additional measurements from stack B and the right-side region of stack A are presented in Supplemental Material, also show similar results. We use a low-frequency ac lock-in detection method to measure the charge and spin transport properties of the graphene flake. In order to measure the I-V behavior of the bottom WS$_2$ flake and for gate-voltage application, a Keithley 2410 dc voltage source was used. All measurements are performed at Helium temperature (4 K) under vacuum conditions in a cryostat.
  
Details of charge and spin-transport measurement methods and TMD characterization are provided in Supplemental Material.  We obtain the BLG electron-mobility $\mu_{\text{e}}\sim$ 3,000 cm$^2$V$^{-1}$s$^{-1}$, which is somewhat low compared to the previously reported mobility values in graphene on a TMD substrate \cite{omar_spin_2018,wang_origin_2016}. 

\begin{figure}
\includegraphics[scale=1]{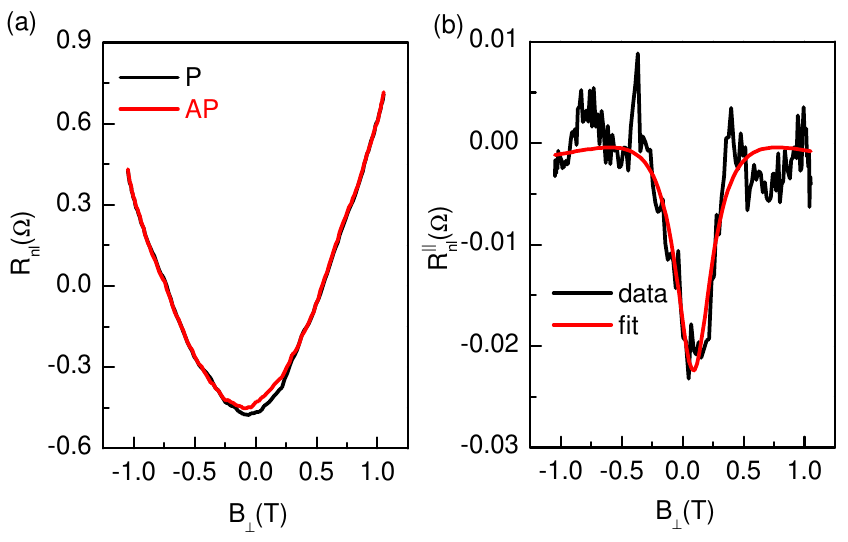}
\caption{\label{fig2} (a) Parallel (P) and anti-parallel (AP) Hanle curves for $L$= 1$\mu$m ($V_{\text{bg}} = 0 V$) show a strong increase in the nonlocal resistance with the applied out-of-plane magnetic field $B_{\perp}$, which indicates a large spin-relaxation anisotropy and the high spin-relaxation time for the out-of-plane spins.  Signs of P and AP configurations are reversed because one electrode has a negative  contact-polarization for in-plane spins. (b) The Hanle spin-signal $R_{\text{nl}}^{||}$ and the fit  result in low $\tau_{||}\sim$ 30ps (stack A). }
\end{figure}

We perform spin-transport measurements, using the measurement scheme shown in Fig.~\ref{fig1}(a) and measure the nonlocal signal $R_{\text{nl}}=v_{\text{nl}}/i_{\text{ac}}$. For in-plane spin transport, the spin-signal is defined as $R_{\text{nl}}^{||}$=$\frac{R_{\text{nl}}^{\text{P}}-R_{\text{nl}}^{\text{AP}}}{2}$, where $R_{\text{nl}}^{\text{AP(P)}}$ is the $R_{\text{nl}}$ measured for the (anti-)parallel magnetization orientations of the injector-detector electrodes. From  non-local spin-valve (SV) and Hanle spin-precession measurements, we obtain the spin diffusion coefficient $D_{\text{s}}$ and in-plane spin-relaxation time $\tau_{||}$, and estimate the spin-relaxation  length $\lambda_{\text{s}}^{||}$= $ \sqrt{D_{\text{s}}\tau_{||}}$.  A representative Hanle measurement for stack A is shown in Fig.~\ref{fig2}(b). Due to small magnitudes of in-plane spin-signals and invasive ferromagnetic (FM) contacts ($\sim$ 1k$\Omega$), we were able to get information about the in-plane spins via Hanle measurements only for short injector-detector separation of about 1-2 $\mu$m. Since we could not access the hole-doped regime for the applied back-gate voltage due to heavily n-doped samples, we only measure the spin-transport in the electron-doped regime for both stacks. For stack A, we obtain $D_{\text{s}}\geq$ 0.01 m$^2$s$^{-1}$ and $\tau_{||}$ in the range 18-34 ps, i.e. $\lambda_{\text{s}}^{||} \sim$ 0.45-0.54 $\mu$m. For stack B, we obtain $D_{\text{s}}\sim$ 0.03 m$^2$s$^{-1}$ and $\tau_{||}$ in the range 17-24 ps, i.e. $\lambda_{\text{s}}^{||} \sim$ 0.6-0.7 $\mu$m. In conclusion, though for both samples we obtain reasonable charge transport properties, i.e. $D_{\text{s}}\sim$ 0.01 m$^2$s$^{-1}$, we obtain a very low $\tau_{||}$ down to 16 ps. The weak modulation of $\tau_{||}$ with the back-gate voltage suggests a strong SOC induced in the BLG in contact with WS$_{\text{2}}$ \cite{wang_origin_2016} and the insignificant contribution of the spin-absorption mechanism for the applied back-gate voltage range in contrast with the behavior observed in refs. \cite{yan_two-dimensional_2016, dankert_electrical_2017, benitez_strongly_2018}.

In order to explore the proposed spin-relaxation anisotropy in BLG/WS$_{\text{2}}$ systems \cite{gmitra_proximity_2017}, we inject out-of-plane spins electrically by controlling the magnetization direction of the FM electrodes via an external magnetic field. Due to its finite shape anisotropy along the z-axis, the magnetization of the FM electrode does not stay in the device plane at high enough $B_{\perp}$. For the FM electrodes with the thickness $\sim$ 65 nm, their magnetization can be aligned fully in the out-of-plane direction at $B_{\perp}\sim$ 1.5 T \cite{tombros_anisotropic_2008,leutenantsmeyer_observation_2018}. At $B_{\perp} \geq$ 0.3 T, the magnetization makes an angle $\theta >$ 10$^{\circ}$ with the easy-axis of the FM electrode, which increases with the field (see Supplemental Material for details). In this case, the injected spins, along with the dephasing in-plane spin-signal component as shown in Fig.~\ref{fig2}(b) also have a non-precessing out-of-plane spin-signal component, which would increase with $B_{\perp}$ due to the contact magnetization aligning towards $B_{\perp}$ (Fig.~\ref{fig2}(a)). From this measurement, we can estimate $\tau_{\perp}$ by removing the contribution of the in-plane spin-signal and the background charge (magneto)resistance, i.e. $R_{\text{sq}}(B_{\perp})$ (for details, refer to Supplemental Material) and fit $R_{\text{nl}}$ with the following equation:
\begin{equation}
 R_{\text{nl}}(B_\perp)=\frac{P^2 R_{\text{sq}}\lambda_{\text{s}}^\perp e^{-\frac{L}{\lambda_{\text{s}}^\perp}}(\sin \theta)^2}{2w}.
 \label{eq1}
\end{equation}
Here $R_{\text{nl}}(B_\perp)$ is the measured signal for out-of-plane spins for the injector-detector separation $L$, channel width $w$, with out-of-plane spin relaxation length $\lambda_{\text{s}}^\perp$. $R_{\text{sq}}$ is the graphene sheet resistance at $B_{\perp}=0$ T. We assume that both electrodes have equal spin-injection and detection polarization $P$, which we obtain in the range 3-5$\%$ via regular in-plane spin-transport measurements (see Supplemental Material for details).
 \begin{figure}
\includegraphics[scale=1]{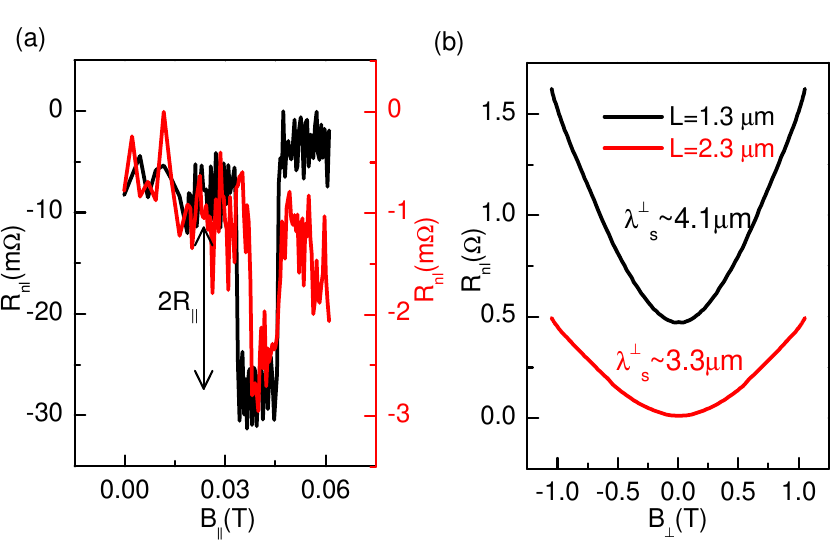}
\caption{\label{fig2.1} (a) In-plane SV signals at the injector-detector separation $L$ =1.3 $\mu$m (black) and 2.3 $\mu$m (red) with their values on the left and right axis, respectively. A background signal of 0.5$\Omega$( 7m$\Omega$) has been subtracted from the measured spin-signal at $L$=1.3(2.3) $\mu$m for a clear representation. (b) Measured and symmetrized Hanle curves for different $L$ for the parallel configuration of FM electrodes.}
\end{figure}

BLG on TMD is expected to have $\tau_{\perp}>>\tau_{||}$ \cite{gmitra_proximity_2017}, which also implies that $R_{\text{nl}}(B_\perp)$ at $\theta=\pi/2$, i.e. $R_{\text{nl}}^\perp$ will be higher in magnitude than $R_{\text{nl}}^{||}$ at $B_{\perp} = 0$. In our measurements, this effect reflects as a strong increase in $R_{\text{nl}}$ at high $B_{\perp}$ for both P and AP configurations (Fig.~\ref{fig2}(a)). 
Via charge magnetoresistance measurements (see Supplemental Material) for the same channel, we confirm that the observed enhancement in $R_{\text{nl}}$ is not due to the  magnetoresistance originating from the orbital effects under the applied out-of-plane magnetic field. Next, we show the distance dependence of $R_{\text{nl}}$ in Fig.~\ref{fig2.1}. The in-plane spin signal $R_{\text{nl}}^{||}$ is reduced almost by factor of 
ten from 10 m$\Omega$ to 1 m$\Omega$ (Fig.~\ref{fig2.1}(a)). On the other hand, $R_{\text{nl}}(B_{\perp})$ for the same distance decreases roughly by less than factor of three. From this measurement, we confirm that  $\tau_{\perp} >> \tau_{||}$ in the BLG/WS$_{\text{2}}$ heterostructures. We fit the experimental data in Fig.~\ref{fig2.1}(b) with Eq.~\ref{eq1} for different $L$,  and obtain $\lambda_{\text{s}}^{\perp} \sim$ 3.3 $\mu$m-4.1 $\mu$m. We extract  $\tau_{\perp}$ from the relation $\lambda_{\text{s}}^{\perp}=\sqrt{D_{\text{s}}\tau_{\perp}}$, while we assume equal $D_{s}$ for in-plane and out-of-plane spins \cite{benitez_strongly_2018}, and obtain $\tau_{\perp}\sim$ 1 ns-1.6 ns, resulting in a large anisotropy $\eta \sim$ 50-70. 
% Note that this method may sometime result in the overestimation of $\eta$ due to the presence of multiple effects \cite{safeer_room-temperature_2019, garcia_spin_2017,huang_anomalous_2017} that create charge background via spin-to-charge conversion, and provides an upper bound on $\eta$ (see Supplemental Material for additional measurements). 

In order to confirm the spin life-time anisotropy in BLG/WS$_{\text{2}}$ system and to accurately measure the out-of-plane spin-signals even in the possible presence of a background charge-signal, we develop a new tool; \textit{Oblique Spin-Valve} (OSV) measurements. For the OSV measurements, we follow a similar measurement procedure as in the SV measurements. However, for the magnetization reversal of FM electrodes, we apply a magnetic field $B$ which makes an angle $\theta_{\text{B}}$ with their easy-axes  in the y-z plane as shown in Fig.~\ref{osv}(a), instead of applying $B_{||}$ in SV measurements  in Fig.~\ref{fig1}(a). As a result, the magnetization of the FM electrodes also makes a finite angle $\theta$ with its easy axis. In this way, we inject and detect both in-plane and out-of-plane spins in the spin-transport channel. The in-plane magnetic field $\sim B\cos\theta_{\text{B}}$ is responsible for the magnetization switching of C2 and C3 (see details in Supplemental Material). At the event of magnetization reversal at a magnetic field in the OSV measurements, the spin-signal change would appear as a sharp switch in $R_{\text{nl}}$. However, the magnetic field dependent background signal does not change.  In this way, in the OSV measurements, we combine the advantages of both SV and the perpendicular-field Hanle measurements, and obtain background-free pure spin-signals.  
\begin{figure}
\includegraphics[scale=0.95]{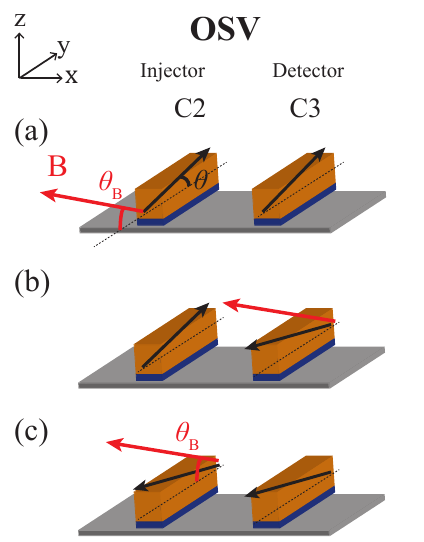}
\includegraphics[scale=1.1]{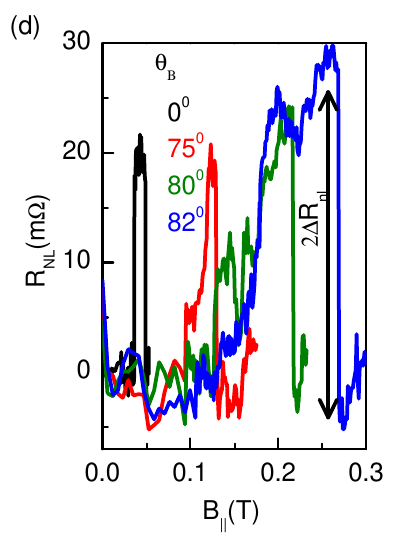}
\caption{\label{osv} (a)-(c) Steps for  Oblique Spin-valve (OSV) measurements. The magnetization vector for the injector and detector (in black) makes an angle $\theta$ with the easy axis and the applied magnetic field $B$ (red vector) for the magnetization reversal remains fixed at an angle $\theta_{\text{B}}$ throughout the measurement. The magnetization reversal for the detector and the injector are shown in (b) and (c), respectively. (d) OSV measurements at different $\theta_{\text{B}}$ values for the injector-detector separation $L$=1 $\mu$m. The OSV spin-signal $\Delta R_{\text{nl}}$ is defined as half of the magnitude of the switch, labeled with the black arrow. The increase in the spin-valve signal magnitude at higher $\theta_{\text{B}}$ confirms the presence of a large spin-relaxation anisotropy.  A background signal ($\sim$0.5-1$\Omega$) has been removed from the measured signal for a clear representation (see Supplemental Material for the original measurement).}
\end{figure}

In an OSV measurement, we measure fractions of both $R_{\text{nl}}^{||}$ and $R_{\text{nl}}^{\perp}$. The OSV spin-signal $\Delta R_{\text{nl}}$ consists of two components: an in-plane spin-signal component proportional to $R_{\text{nl}}^{||}\cos^2\theta$  and an out-of-plane spin-signal component proportional to  $R_{\text{nl}}^{\perp}\sin^2\theta$ which get dephased by the applied magnetic field $B\sin\theta_{\text{B}}$ and $B\cos\theta_{\text{B}}$, respectively:

\begin{equation}
  \Delta R_{\text{nl}} \simeq  R_{\text{nl}}^{||}\cos^2\theta\zeta_{||}({B\sin\theta_{\text{B}}}) +  R_{\text{nl}}^{\perp}\sin^2\theta\zeta_{\perp}({B\cos\theta_{\text{B}}})
  \label{sv osv}
 \end{equation}
where $\zeta_{||(\perp)}$ is the functional form for the in-plane (out-of-plane) spin precession dynamics.  At larger $\theta_{\text{B}}$, the dephasing of in-plane spin-signal $R_{\text{nl}}^{||}$ is enhanced. Conversely, the dephasing of out-of-plane spin-signal$R_{\text{nl}}^{\perp}$ is suppressed. Also, $\theta$ increases with $\theta_{\text{B}}$. Therefore, $\Delta R_{\text{nl}}$ at higher $\theta_{\text{B}}$ is dominated by $R_{\text{nl}}^{\perp}$ and acquires a similar form as in Eq.~\ref{eq1}. 

Due to the expected spin-life time anisotropy in BLG/TMD systems and as observed in Hanle measurements in Fig.~\ref{fig2.1}(b), the out-of-plane spin signal magnitude increases with the magnetization angle $\theta$. Similar effect would appear in the OSV measurements at larger $\theta_{\text{B}}$ values due to fact that the magnetization switching would occur at larger $\theta$, which would allow to measure a larger fraction of the out-of-plane spin-signal. In order to verify our hypothesis, we first measure the in-plane spin-valve signal $\Delta R_{\text{nl}}=R_{\text{nl}}^{||}$ at $\theta_{\text{B}}$ = 0$^\circ$  for L=1 $\mu$m, and then measure $R_{\text{nl}}$ at different $\theta_{\text{B}}$ values. The measurement summary is presented in Fig.~\ref{osv}(d). Here, we clearly observe an  increase in $\Delta R_{\text{nl}}$  up to 1.5 times with the increasing $\theta_{\text{B}}$.  This result is remarkable in the way that it is possible to observe such clear enhancement even with a small fraction of $R_{\text{nl}}^{\perp}$, i.e. $\propto R_{\text{nl}}^{\perp}\sin^2\theta $ contributing  to $\Delta R_{\text{nl}}$. Note that, following Eq.~\ref{sv osv}, for $\eta \leq$ 1 (or $R_{\text{nl}}^{\perp}\leq R_{\text{nl}}^{||}$), we would never observe an increase in $R_{\text{nl}}$. Therefore the observation of an enhanced signal in the OSV measurements is the confirmation of the present large spin life-time anisotropy in the BLG/WS$_{\text{2}}$ system.

In order to simplify the analysis and to estimate $R_{\text{nl}}^{\perp}$ from the OSV measurements, we assume that the out-of-plane signal is not significantly affected by the in-plane magnetic field component ($\sim$ 10 mT) at $\theta_{\text{B}} >$ 80$^{\circ}$,  and $\zeta_{\perp}(B\cos\theta_{\text{B}})$ can be omitted from Eq.~\ref{sv osv}. Note that this assumption would lead to the lower bound of $R_{\text{nl}}^{\perp}$ or $\tau_{\perp}$.   $R_{\text{nl}}^{||}$ and $\zeta_{||}$ are obtained via the in-plane SV and Hanle spin-precession measurements (for details refer to Supplemental Material). From $R_{\text{nl}}^{\perp}$, we obtain $\lambda_{\text{s}}^{\perp}\sim$ 3.7-4 $\mu$m, which is similar to $\lambda_{\text{s}}^{\perp}$ obtained via Hanle measurements, and confirms the validity of the analysis. 
Using $\lambda_{\text{s}}^{\perp}=\sqrt{D_{\text{s}}\tau_{\perp}}$, we estimate $\tau_{\perp} \sim$ 1-2 ns and the lower limit of $\eta \sim$ 70 for $V_{\text{bg}}$ between -45 V to 40 V except at $V_{\text{bg}} =$ -20V (Fig.~\ref{fig3}(a)). Such high magnitude of $\tau_{\perp} \sim$ ns is also expected theoretically even in presence of spin-orbit coupling \cite{gmitra_proximity_2017}, is comparable to the spin relaxation times observed in ultra-clean graphene \cite{gurram_bias_2017, ingla-aynes_$24ensuremath-ensuremathmumathrmm$_2015,drogeler_spin_2016,leutenantsmeyer_observation_2018}, and is a clear signature of strong spin-valley coupling present in the BLG/WS$_{\text{2}}$ system (see Supplemental Material for additional measurements).
\begin{figure}
\includegraphics[scale=1]{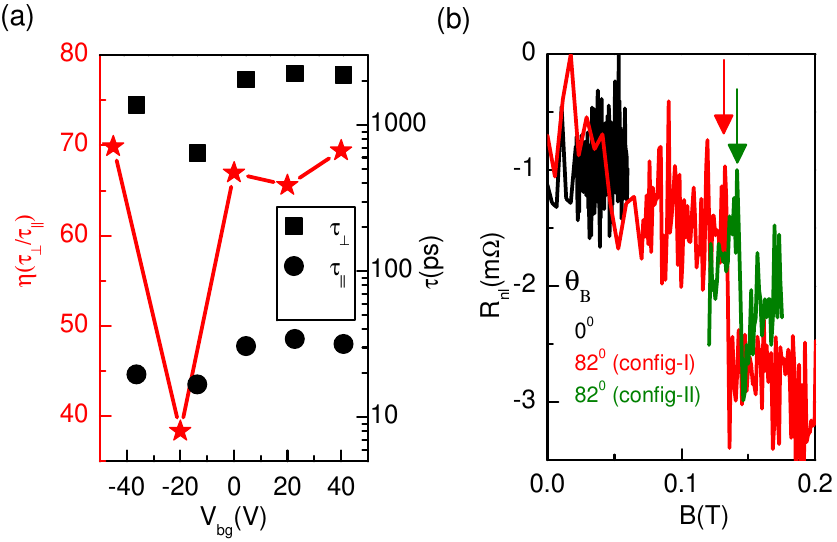}
\caption{\label{fig3}(a) $\eta-V_{\text{bg}}$ plot (in red) on the left y-axis, and respective $\tau_{||}$ and $\tau_{\perp}$ as a function $V_{\text{bg}}$ on the right y-axis(b) OSV measurements at $L$=4.3 $\mu$m at $\theta_{\text{B}} =$ 0$^{\circ}$(black curve), $\theta_{\text{B}} =$ 82$^{\circ}$ (config.-I) and for the swapped injector-detector (config.-II) at $\theta_{\text{B}} =$ 82$^{\circ}$. Arrows in the figure indicate the switching of electrode $C2$ in Fig.~\ref{osv} }
\end{figure}

In presence of  large $\eta$ values in BLG/WS$_{\text{2}}$ heterostructures, the out-of-plane spin-signal can still be detected at larger distances via OSV measurements whereas the in-plane is not even possible to detect. We present such a case in  Fig.~\ref{fig3}(b) for L = 4.3 $\mu$m, where no in-plane spin-signal is detected. However, we clearly measure $\Delta R_{\text{nl}}$ = 1.5 m$\Omega$ for $\theta_{\text{B}} =$ 82$^{\circ}$, and obtain a similar result by swapping the injector and detector electrodes. The presented measurement unambiguously establishes the fact that indeed due to extremely large $\eta$, even though we measure a small fraction $\sim R_{\text{nl}}^{\perp}\sin^2\theta$ of $R_{\text{nl}}^{\perp}$, its magnitude is larger than the in-plane spin-signal.  

In summary, we report the first spin-transport measurements on a bilayer-graphene/TMD system. We find low in-plane spin relaxation times in the range of 20-40 ps which weakly depend on the carrier density and conductivity of the underlying TMD, and therefore suggest a strong proximity induced spin-orbit coupling in the BLG. Via Hanle and OSV measurements, we electrically inject and detect out-of-plane spins in the BLG/WS$_{\text{2}}$ system. We  estimate the out-of-plane spin relaxation time $\sim$ 1-2 ns and the anisotropy value between 40$\sim$70. It is noteworthy that obtained $\eta$ and $\tau_{\perp}$ for BLG/TMD are much larger compared to previously reported values in Gr/TMD systems in refs. \cite{ghiasi_large_2017,benitez_strongly_2018}.  These results confirm the theoretical prediction that the BLG/TMD systems are highly anisotropic, and show efficient spin-valley coupling for out-of-plane spins. Obtained results unlock the potential of single layer graphene/TMD systems and would be crucial in developing future spintronic devices such as efficient spin-filters.

% with the Larmor frequency $\overrightarrow{\omega_{\text{L}}}= \frac{g \mu_{\text{B}}}{\hbar}{B_{\perp}}$,  while diffusing towards the detector, and gets dephased. Here $g$ is the gyromagnetic ratio(=2) for an electron, $\mu_{\text{B}}$ is the Bohr magneton and $\hbar$ is the reduced Planck constant.  The measured Hanle curves are fitted with the steady state solution to the one-dimensional Bloch equation \cite{tombros_electronic_2007}:
% \begin{equation}
%  D_{\text{s}} {\bigtriangledown}^2\overrightarrow{\mu_{\text{s}}}-\frac{\overrightarrow{\mu_{\text{s}}}}{\tau}+\overrightarrow{\omega_{\text{L}}}\times \overrightarrow{\mu_{\text{s}}}=0
% \label{bloch}
% \end{equation}

% \section{Discussion}
 
%  \section{conclusions}
%  \section{Acknowledgements}
We acknowledge  J. G. Holstein, H.H. de Vries, T. Schouten and H. Adema for their technical assistance. We thank M.H.D. Guimar\~{a}es for critically reading the manuscript. This research work was funded by the the Graphene flagship core 1 and core 2 program  (grant no. 696656 and 785219), Spinoza Prize (for B.J.v.W.) by the Netherlands Organization for Scientific Research (NWO) and supported by the Zernike Institute for Advanced Materials.

%  \bibliography{WS2_BLG_paper10.bib}

\externaldocument{paper_WS2_BLG_final}

\hypersetup{bookmarksnumbered, pdfpagemode=UseOutlines, 
colorlinks=true, citecolor=blue, filecolor=blue, linkcolor=blue, urlcolor=blue}

%\usepackage[mathlines]{lineno}% Enable numbering of text and display math
%\linenumbers\relax % Commence numbering lines

%\usepackage[showframe,%Uncomment any one of the following lines to test 
%%scale=0.7, marginratio={1:1, 2:3}, ignoreall,% default settings
%%text={7in,10in},centering,
%%margin=1.5in,
%%total={6.5in,8.75in}, top=1.2in, left=0.9in, includefoot,
%%height=10in,a5paper,hmargin={3cm,0.8in},
%]{geometry}
% \graphicspath{/media/sid/Dropbox/2nd_paper
% }
%\begin{document}

\beginsupplement

\newpage
\begin{center}
 \textbf{\large Supplementary Information}
\end{center}

\section{sample preparation}
Tungsten disulfide (WS$_2$) flakes are exfoliated on a polydimethylsiloxane (PDMS) stamp and identified using an optical microscope. The desired flake is transferred onto a pre-cleaned SiO$_2$/Si substrate ($t_{\text{SiO$_2$}}$=500 nm), using a transfer-stage. The transferred flake on SiO$_2$ is annealed in an Ar-H$_2$ environment at 240$^{\circ}$C for 6 hours in order to achieve a clean top-interface of WS$_2$, to be contacted with graphene. The graphene flake is exfoliated from a ZYB grade HOPG (Highly oriented pyrolytic graphite) crystal and boron nitride (BN) is exfoliated from BN crystals (size$\sim$ 1 mm) onto different SiO$_2$/Si substrates ($t_{\text{SiO$_2$}}$=90 nm). Both crystals were obtained from HQ Graphene. The desired bilayer-graphene (BLG) flakes are identified via their optical contrast using an optical microscope. Boron-nitride flakes are identified via the optical microscope. The thickness of hBN and WS$_{\text{2}}$ flakes is determined via Atomic Force Microscopy. In order to prepare an hBN/Gr/WS$_2$ stack, we use a polycarbonate (PC) film attached to a PDMS stamp as a sacrificial layer. Finally, the stack is annealed again in the Ar-H$_2$ environment for six hours at 235$^{\circ}$C to remove the remaining PC polymer residues. 

In order to define contacts, a poly-methyl methacrylate (PMMA) solution is spin-coated over the stack and the contacts are defined via the electron-beam lithography (EBL). The PMMA polymer exposed via the electron beam gets dissolved in a MIBK:IPA (1:3) solution. In the next step, 0.7 nm Al is deposited in two steps, each step of 0.35 nm followed by 12 minutes oxidation in the oxygen rich environment to form a AlO$_x$ tunnel barrier. On top of it, 65 nm thick cobalt (Co) is deposited to form the ferromagnetic (FM) tunnel contacts with a 3 nm thick Al capping layer to prevent the oxidation of Co electrodes. The residual metal on the polymer is removed by the lift-off process in acetone solution at 40$^{\circ}$C. 

% \section{Sample Characterization}

%  Here, we use top-hBN for independent control of charge carriers and electric-field in the encapsulated-BLG region. We use the low-frequency lock-in detection method to measure the charge and spin transport properties of the graphene flake. In order to measure the I-V behavior of the WS$_2$ flake and for gate-voltage application, a Keithley 2410 dc source meter was used. All measurements are performed at Helium temperature (4 K) under vacuum conditions in a cryostat.
 
\section{Charge transport measurements}  
\subsection{Graphene}
\begin{figure*}[!t]
\includegraphics[scale=2]{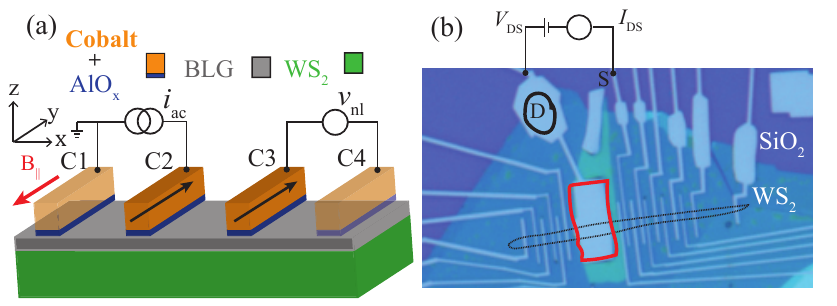}
\caption{\label{S1} (a) Nonlocal spin-transport measurement scheme. (b) An optical micrograph of a fabricated WS$_{\text{2}}$/BLG/hBN stack (stack A). BLG is outlined with black dashed lines and hBN top-gate is outlined in red.}
\end{figure*}
\begin{figure}[!t]
\includegraphics[scale=1]{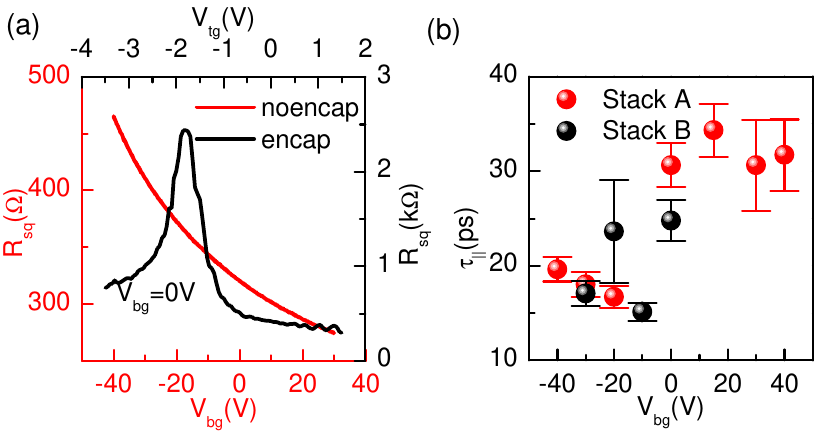}
\caption{\label{S2} (a)$R_{\text{sq}}-V_{\text{bg(tg)}}$ dependence for the nonencapsulated (encapsulated) region is shown on the left (right) axis (red(black) curves) for stack A. (b) $\tau_{||}-V_{\text{bg}}$ for BLG/WS$_{\text{2}}$}
\end{figure}

We measure the charge transport in graphene via the four-probe local measurement scheme. For measuring the gate-dependent resistance of graphene-on-WS$_2$, a fixed ac current $i_{\text{ac}} \sim$ 100 nA is applied between contacts C1-C4 and the voltage-drop is measured between contacts C2-C3 (Fig.~\ref{S1}(a)), while the back-gate voltage is swept. The maximum resistance point in the Dirac curve is denoted as the charge neutrality point (CNP). For graphene-on-WS$_2$, it is possible to tune the Fermi energy $E_{\text{F}}$ and the carrier-density in graphene only when $E_{\text{F}}$ lies only in the band-gap of WS$_2$. Since, we do not observe any saturation in the resistance of the BLG (red curve Fig.~\ref{S2}(a)), we probe the charge/ spin transport where the Fermi level lies within the band gap of WS$_2$. The CNP cannot be accessed within the applied $V_{\text{bg}}$ range. However, it is possible to access the CNP and the hole doped regime (black curves Fig.~\ref{S2}(a)) in the region underneath the top-hBN flake, outlined as red region in the optical image in Fig.~\ref{S1}(b), using the top-gate application due to its higher capacitance.
In order to extract the carrier mobility $\mu$, we fit the charge-conductivity $\sigma$ versus carrier density $n$ plot with the following equation:
\begin{equation}
 \sigma=\frac{1}{R_{\text{sq}}}=\frac{ne\mu+\sigma_0}{1+R_{\text{s}}(ne\mu+\sigma_0)}.
 \label{mobility}
\end{equation}
 Here $R_{\text{sq}}$ is the square resistance of graphene, $\sigma_0$ is the conductivity at the CNP, $R_{\text{s}}$ is the residual resistance due to short-range scattering \cite{gurram_spin_2016,zomer_fast_2014} and $e$ is the electronic charge. We fit the $\sigma-n$ data for $n$ (both electrons and holes) in the range 0.5-2.5$\times$10$^{12}$ cm$^{-2}$ with Eq.~\ref{mobility}. For the encapsulated region we obtain the electron-mobility $\mu_{\text{e}}\sim$ 3,000 cm$^2$V$^{-1}$s$^{-1}$ for stack A. For stack B, we could not access the CNP within the applied $V_{\text{bg}}$ range due to heavily n-doped BLG. Therefore, we could not extract the mobility. 
 \subsection{Tungsten disulfide (WS$_2$)}
\begin{figure}[!t]
\includegraphics[scale=1]{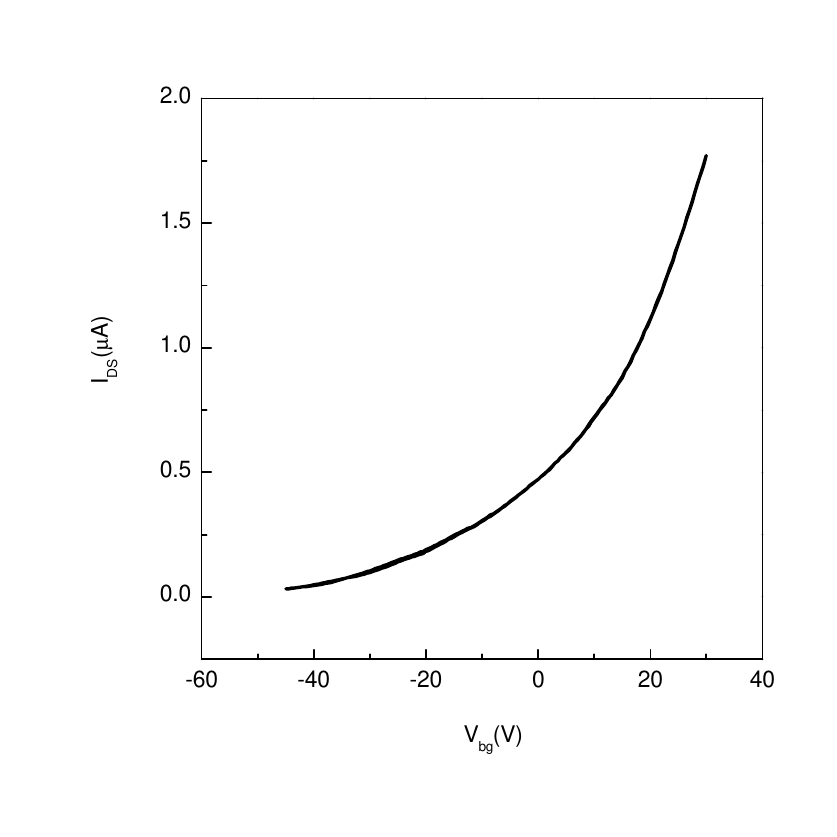}
\caption{\label{Tr} $I_{\text{DS}}-V_{\text{bg}}$ behavior of the bottom-WS$_2$ flake of stack A at $V_{\text{DS}}$ =0.2 V applied between the top-gate electrode and another electrode contacting the BLG-on-WS$_2$. The measurement scheme is shown in Fig.~\ref{S1}(b).}
\end{figure}

 In order to obtain the transfer characteristics, i.e. back-gate dependent conductivity of the WS$_2$ substrate, we apply a dc voltage $V_{\text{DS}}$ = 0.2 V and measured the current $I_{\text{DS}}$ between the top gate contact, that touches the bottom WS$_2$ at point D and a contact S on the BLG flake (Fig.~\ref{S1}(b)), and vary the back-gate voltage $V_{\text{bg}}$ in order to change the resistivity of WS$_2$. The $I_{\text{DS}}-V_{\text{bg}}$ behavior of the bottom-WS$_2$ flake of stack A is plotted in Fig.~\ref{Tr}.
 \section{Spin transport measurements} 
 For spin-valve (SV) measurements, a charge current $i_{\text{ac}}$ is applied between contacts C2-C1 and a nonlocal voltage $v_{\text{nl}}$ is measured between C3-C4 (Fig.~\ref{S1}(a)). First an in-plane magnetic field $B_{||}\sim$ 0.2 T is applied along the easy axes of the ferromagnetic (FM) electrodes (+y-axis), in order to align their magnetization along the field. Now, $B_{||}$ is swept in the opposite direction (-y-axis) and the FM contacts reverse their magnetization direction along the applied field, one at a time. This magnetization reversal appears as a sharp transition in $v_{\text{nl}}$ or in the nonlocal resistance $R_{\text{nl}}=v_{\text{nl}}/i_{\text{ac}}$. The spin-signal is $ R_{\text{nl}}^{||}=\frac{R_{\text{nl}}^{\text{P}}-R_{\text{nl}}^{\text{AP}}}{2}$, where $R_{\text{nl}}^{\text{P(AP)}}$ represents the $R_{\text{nl}}$ value of the two level spin-valve signal, corresponding to the parallel (P) and anti-parallel (AP) magnetization of the FM electrodes. In the nonlocal measurement geometry the spin-signal $R_{\text{nl}}^{||}$ is given by:
 
 \begin{equation}
 R_{\text{nl}}^{||}=\frac{P^2 R_{\text{sq}}\lambda_{\text{s}}^{||} e^{-\frac{L}{\lambda_{\text{s}}^{||}}}}{2w}.
 \label{rnl ip}
\end{equation}

Here $\lambda_{\text{s}}^{||}$ is the spin-relaxation length for the in-plane spins in graphene and $P$ is the contact polarization of injector and detector electrodes for in-plane spins, $R_{\text{sq}}$ is the graphene sheet-resistance and $w$ is the width of spin-transport channel.
  
For Hanle spin-precession measurements, for a fixed P (AP) configuration, an out-of-plane magnetic field $B_{\perp}$ is applied and the injected in-plane spin-accumulation precesses around the applied field. From these measurements, we obtain the spin diffusion coefficient $D_{\text{s}}$ and in-plane spin-relaxation time $\tau_{||}$, and estimate the spin-relaxation  length $\lambda_{\text{s}}^{||}$= $ \sqrt{D_{\text{s}}\tau_{||}}$. Using this $\lambda_{\text{s}}^{||}$ in Eq.~\ref{rnl ip}, we obtain the contact polarization $P\sim$ 3-5 \% for in-plane spin-transport. We would like to make a remark here that some of the contacts in stack A have the opposite (i.e., negative) sign of $P$ for in-plane spin-transport. The origin of the negative sign is nontrivial and possibly could be due to the specific nature of the FM tunnel barrier interface with the graphene-on-TMD.

 \begin{figure}[!t]
\includegraphics[scale=1]{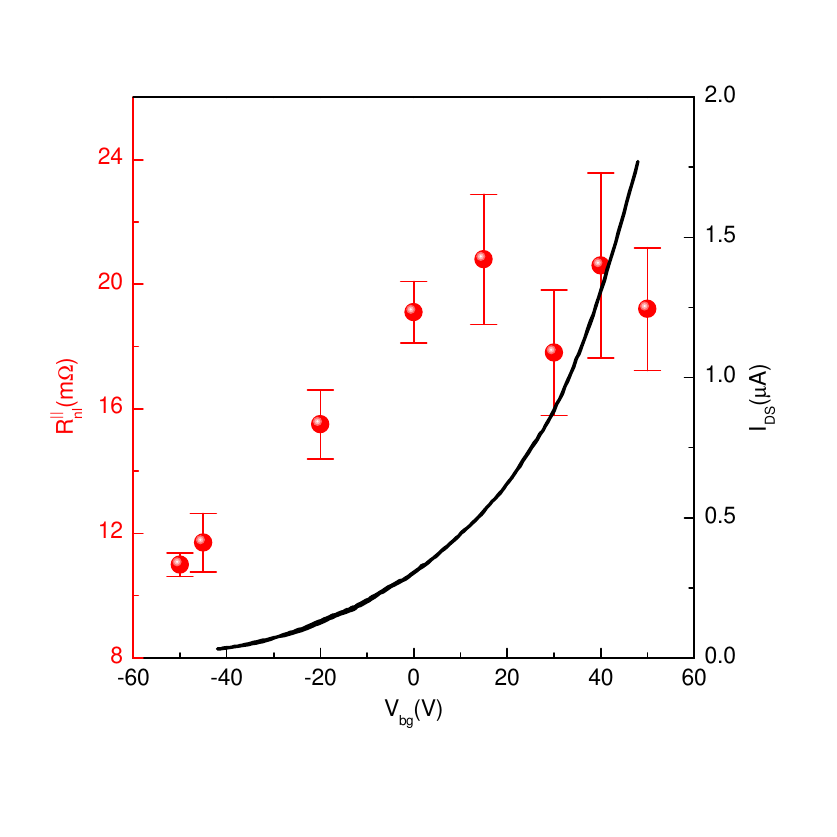}
\caption{\label{sv bg stA} In-plane Spin valve (SV) measurements for stack A as a function of $V_{\text{bg}}$ and the conductance of the underlying TMD (WS$_{\text{2}}$).}
\end{figure}
\begin{figure}[!t]
\includegraphics[scale=1]{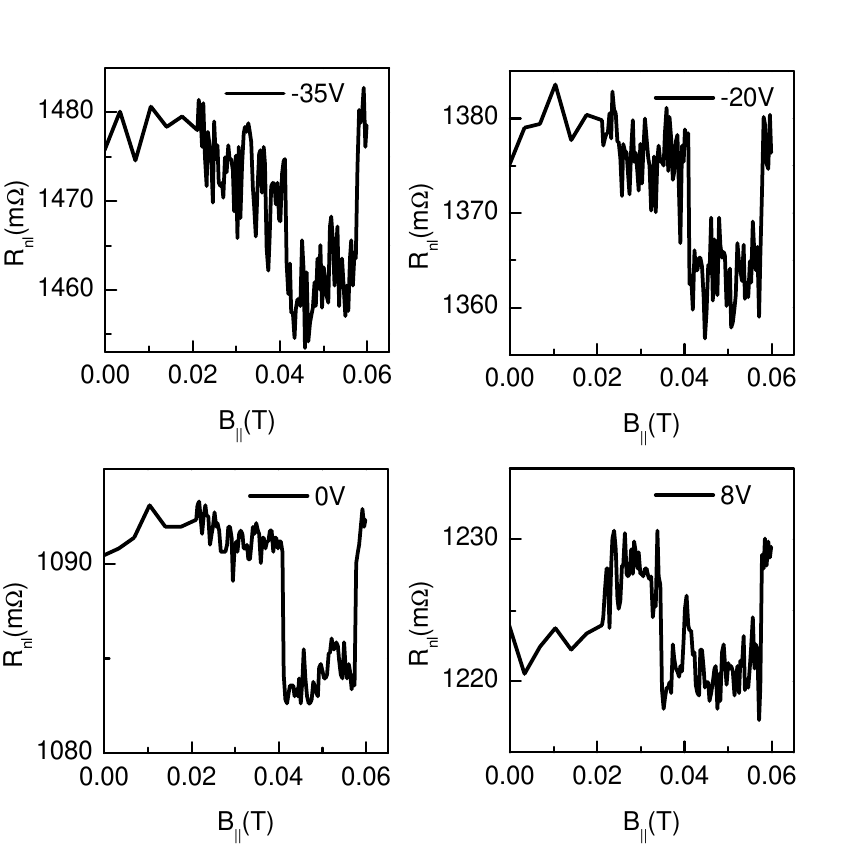}
\caption{\label{sv bg} In-plane Spin valve (SV) measurements for stack B at different back-gate voltage ($V_{\text{bg}}$) values. $R_{\text{nl}}^ {||}$ does not change with $V_{\text{bg}}$, indicating that the spin-absorption is not the dominant mechanism for spin-relaxation within the applied $V_{\text{bg}}$ range.}
\end{figure}

SV measurements as a function of $V_{\text{bg}}$ (stack B) are summarized in Figs.~\ref{sv bg stA} and ~\ref{sv bg} for stack A and stack B, respectively. For both samples, there is no significant change in the spin-signal within the range $\Delta V_{\text{bg}} \sim\pm$ 40V. For stack A, the FM contacts have low resistance ($\leq$ 1k$\Omega$) and this is the reason that there is a modest increase in $R_{\text{nl}}^{||}$ at higher charge carrier density due to the suppressed contact-induced spin-relaxation \cite{maassen_contact-induced_2012,omar_graphene-$mathrmws_2$_2017}. Both measurement do not exhibit any measurable signature of   spin-absorption due to the conductivity modulation of the underlying TMD substrate.

\section{Generalized Stoner-Wohlfarth Model for extracting magnetization angle}
\begin{figure*}[!t]
\includegraphics[scale=1]{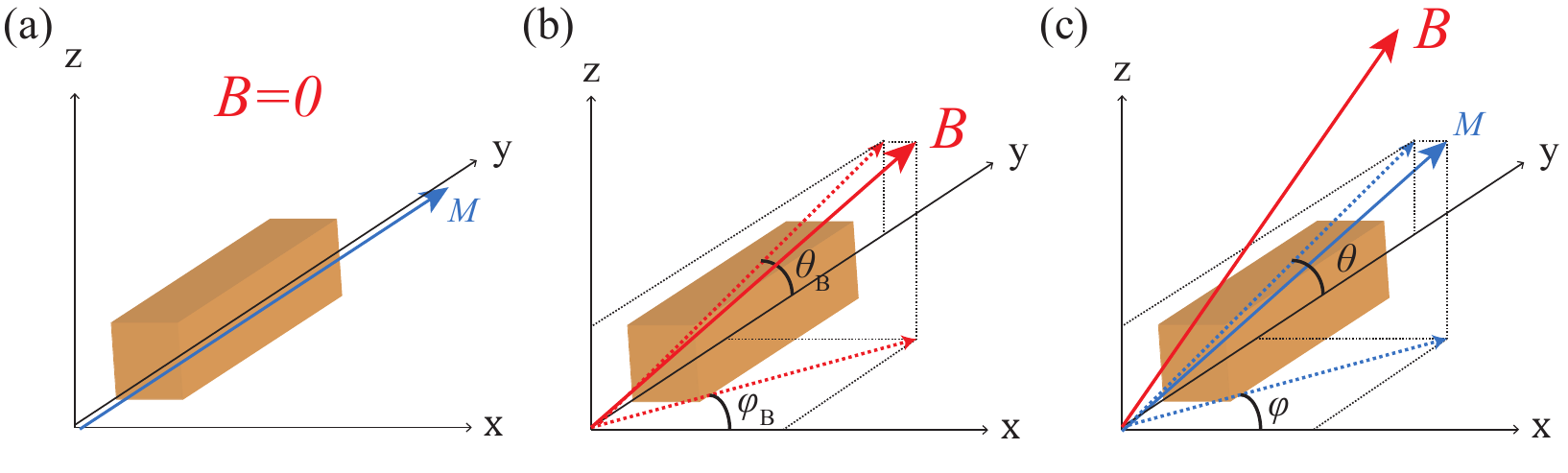}
\caption{\label{mgtc angle} Easy axis of magnetization ${M}$ for the bar magnet is along its length, i.e. along y-axis. (a) $M$ is along y-axis for $B$ = 0 or when $B$ is applied along y-axis. (b) For $B \neq$ 0, $M$ makes an angle $\theta$  with the x-y plane and angle $\phi$ with the y-z plane and (c) $B$ makes an angle $\theta_{\text{B}}$ with the x-y plane and angle $\phi_{\text{B}}$ with the y-z plane.}
\end{figure*}
In this section, we describe the basics of Stoner-Wohlfarth (SW) model, and extend it for three dimensional case in order to extract the magnetization-direction of a bar-magnet in presence of an external magnetic field.

% When a ferromagnet experiences a magnetic field $\overrightarrow{B}$, its magnetization vector $\overrightarrow{M}$ tries to align with $\overrightarrow{B}$ and minimize its potential energy, i.e. also called Zeeman energy $E_{\text{Z}}$ .
% \begin{equation}
%  E_{\text{Z}} = -\overrightarrow{M}.\overrightarrow{B} = -MB\cos\gamma,
%  \label{zeeman}
% \end{equation}
% 
% where $\gamma$ is the angle between $\overrightarrow{M}$ and $\overrightarrow{B}$. However, it is not free to align itself freely as due to the uniaxial anisotropy, it prefers to stay in certain orientation along its magnetization axis. If displaced from this position, it gains the energy in form of anisotropic energy $E_{\text{A}}$:
% 
% \begin{equation}
%  E_{\text{A}} = K \sin^2\alpha,
%  \label{anisotropy}
% \end{equation}
% where $K$ is the uniaxial anisotropic constant and $\alpha$ is the angle between $\overrightarrow{M}$ and the easy axis of magnetization. 

The total energy $E_{\text{T}}$ of a ferromagnet in a magnetic field is expressed as:

\begin{equation}
 E_{\text{T}} = E_{\text{A}} + E_{\text{Z}},
\end{equation}

where $E_{\text{Z}}$ and $E_{\text{A}}$ are the contributions from Zeeman and anisotropic energy, respectively. 

First a magnetic field $B$ is applied which makes an angle $\phi_{\text{B}}$ with the x-axis and an angle $\theta_{\text{B}}$ (Fig.~\ref{mgtc angle}(b)), having its components $B_{x},B_y,B_z$ along $x,y$ and $z$ axes, respectively. Here $B$ can be parameterized with respect to $\theta_{\text{B}},\phi_{\text{B}}$) in the following way:
\begin{equation}
 B_x = B\cos\theta_{\text{B}} \cos\phi_{\text{B}},
 B_y = B\cos\theta_{\text{B}} \sin\phi_{\text{B}},
 B_z = B\sin\theta_{\text{B}}.
\end{equation}

For a ferromagnetic bar with its anisotropic constants $K_x, K_y$ and $K_z$ along x,y and z axis, respectively, and $\overrightarrow{M}$ making an angle $\alpha_{x}, \alpha_{y}$ and $\alpha_{z}$ with the x, y and z axis, respectively, $E_{\text{T}}$ be generalized to a three-dimensional form as:

\begin{equation}
 E_{\text{T}} = \sum_{i=x,y,z}E_{\text{A}}^{i} + \sum_{i=x,y,z}E_{\text{Z}}^{i}
 \label{total energy}
\end{equation}

Now we write down the expression for $E_{\text{A}}^i$ and $E_{\text{Z}}^i$ which have contributions from $M_i$ and $B_i$.

At ($B,\theta_{\text{B}},\phi_{\text{B}}$) $\overrightarrow{M}$ makes the azimuthal angle $\phi$ with the x-axis in the x-y plane and polar angle $\theta$ with the y-axis in the y-z plane (Fig.~\ref{mgtc angle}(c)). Therefore, $\overrightarrow{M}$ = $(M_x, M_y, M_z)$ = $(M\cos\theta \cos\phi, M\cos\theta \sin\phi, M\sin\theta)$. The anisotropic and Zeeman energy terms can again be parameterized with respect to $\theta,\phi,\theta_{\text{B}}, \phi_{\text{B}}$ in to a three-dimensional form:
\begin{align}
\begin{split}
 E_{\text{A}}^x = K_x\sin^2\alpha_x = K_x(1-\cos^2\theta\cos^2\phi),\\ 
 E_{\text{A}}^y = K_y\sin^2\alpha_y = K_y(1-\cos^2\theta\sin^2\phi),\\
 E_{\text{A}}^z = K_z\sin^2\alpha_z = K_z\cos^2\theta,\\
 \label{anisotropy}
\end{split}
 \end{align}
and
\begin{align}
\begin{split}
E_{\text{Z}}^x = -M_xB_x,\\
E_{\text{Z}}^y = -M_yB_y,\\
E_{\text{Z}}^z = -M_zB_z.
\label{zeeman}
\end{split}
\end{align}

Now the expressions in Eq.s \ref{anisotropy} and \ref{zeeman} can be substituted to Eq.~\ref{total energy} and a full functional form of $E_{\text{T}}$ can be obtained. 

In order to obtain $(\theta,\phi)$ which correspond to min($E_{\text{T}}$), we solve for the global energy minima of Eq.~\ref{total energy} by imposing two following conditions:

\begin{figure}[]
\includegraphics[scale=1]{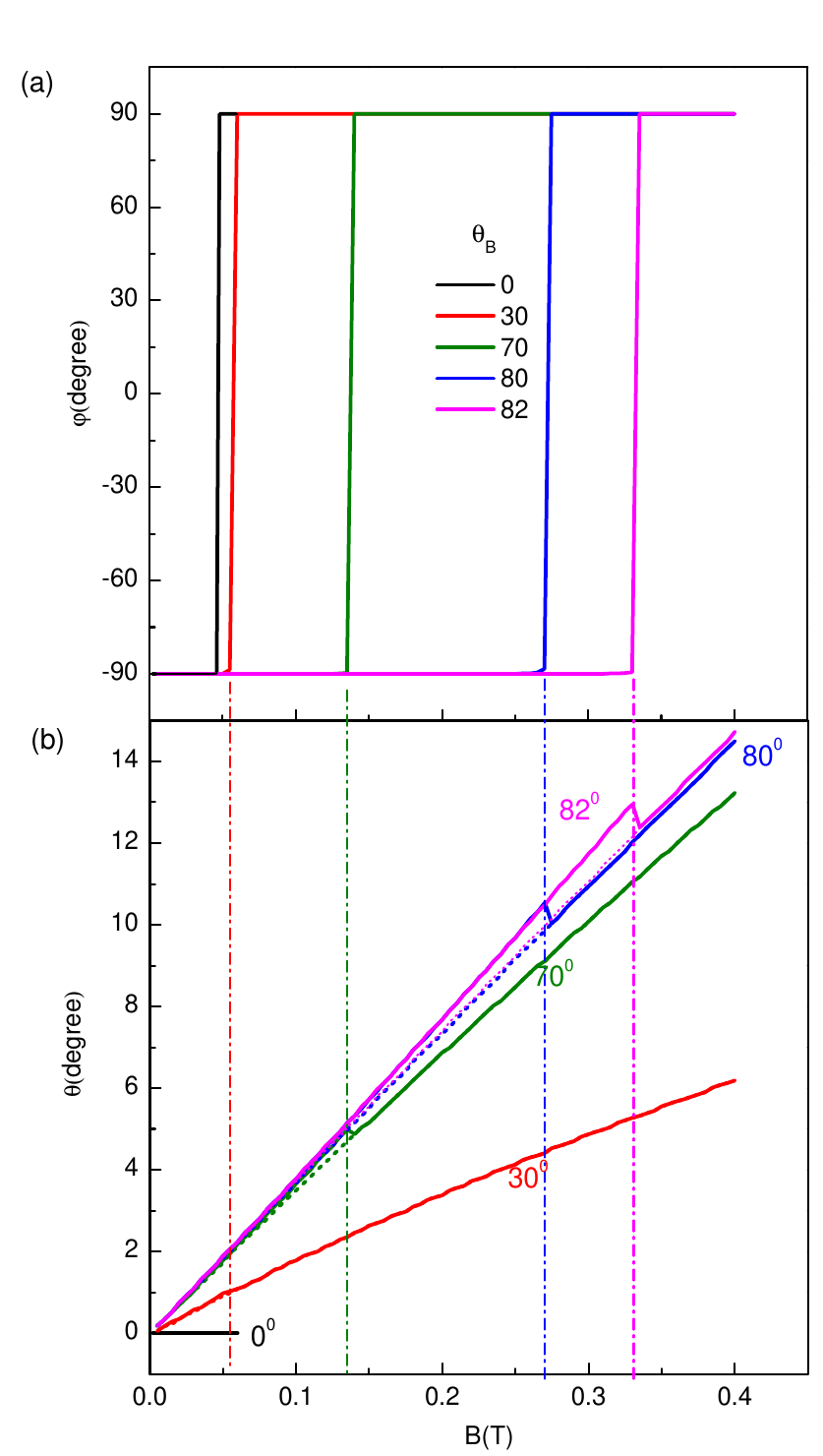}
\caption{\label{simulation} (a) in-plane $\phi$ and (b) out-of-plane $\theta$ angles as a function of magnetic field at different $\theta_{\text{B}}$ values. Dashed lines in the $\theta-B$ plot correspond to the situation when $B$ and $M$ have their in-plane components in the same direction in the y-z plane. Sharp switches in $\phi, \theta$ correspond to the event when the magnetization reversal occurs.}
\end{figure}
\begin{align}
 \begin{split}
  \frac{\partial E_{\text{T}}(\theta,\phi)}{\partial \theta} = \frac{\partial E_{\text{T}}(\theta,\phi)}{\partial \phi} = 0,\\
  \frac{\partial^2 E_{\text{T}}(\theta,\phi)}{\partial \theta^2} = \frac{\partial^2 E_{\text{T}}(\theta,\phi)}{\partial \phi^2} = 0.
  \label{min energy}
 \end{split}
\end{align}

Since $\overrightarrow{M}$ has its easy axis along y-axis, $K_y = 0$. We use $M_{\text{cobalt}}$ = 5$\times$10$^5 A/m$ as reported in literature \cite{kittel_introduction_2004}. In order to obtain $K_z$, we use the saturation magnetic field $M_{\text{s}}$ of the FM electrodes along z-direction, i.e. $\sim$ 1.5 T for the thickness (65nm) of the FM electrodes, and use the relation $M_{\text{s}}=\frac{2K_z}{M_{\text{cobalt}}}$ \cite{raes_determination_2016}. In order to obtain $K_x$, we use the in-plane switching fields of FM electrodes, and use them as the only free parameter in the model to obtain the in-plane magnetization switching as obtained in measurements.  

Using the procedure, we numerically solve for $\theta,\phi$ for different directions of the applied magnetic field with respect to the minimum energy constraint in Eq.~\ref{min energy} using MATLAB. The simulation outcome is shown in Fig.~\ref{simulation}. 

% From the simulation results, we can draw following conclusions:
% 
% 
% \begin{enumerate}
%  \item For the applied field $B$ at an angle $\theta_{\text{B}}$ , in-plane switching occurs expectedly at $B\sim B_0/\cos\theta_{\text{B}}$. Here, $B_0$ is the magnetization switching field for in-plane spin-valve measurements.
% %  \item $\overrightarrow{M}$ subtends a higher angle $\theta$ when $B$ has its in-plane component in the opposite direction of in-plane magnetization component.
%  \item There is a small change in $\theta$ during the magnetization switching at lower $\theta_{\text{B}}$ values, which increases with $\theta_{\text{B}}$.
% \end{enumerate}

\section{Oblique Spin-valve measurements}
\begin{figure}[!h]
\includegraphics[scale=1]{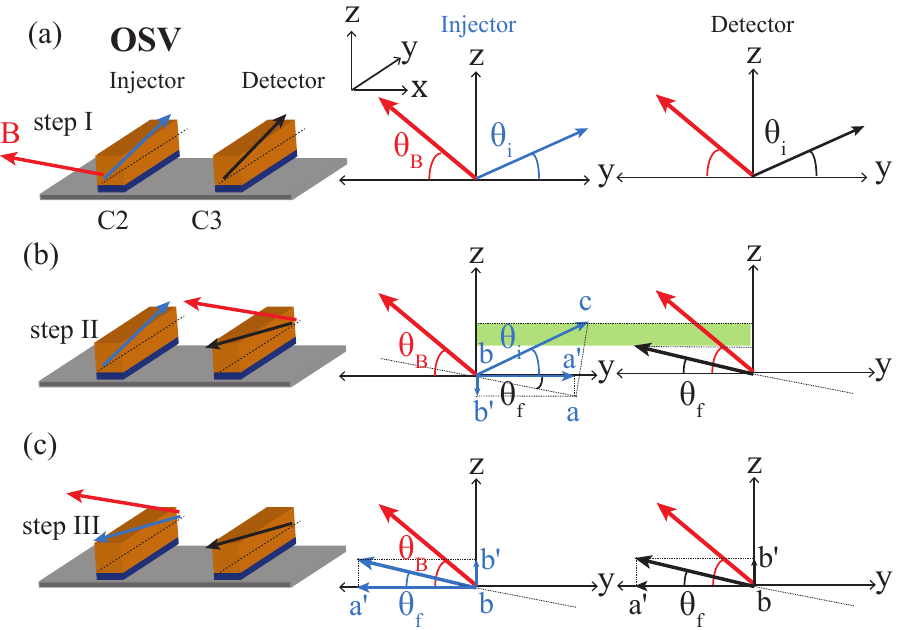}
\caption{\label{osv_steps} Steps for Oblique Spin Valve Measurements}
\end{figure}
Before starting the Oblique Spin-valve (OSV) measurements, we set the initial $\overrightarrow{M}$ of the FM electrodes along +y-axis, i.e. along their easy magnetization-axis. Here, the measured spin-signal $R_{\text{nl}}^{\text{T}} = R_{\text{nl}}^{||}$.

{\bf Step-I:} We apply a magnetic field $B$ in the opposite direction which makes an angle $\theta_{\text{B}}$ with the -y-axis, as shown in Fig.~\ref{osv_steps}(a). Here, we assume that both injector and the detector due to their identical thickness have the same out-of-plane anisotropy  value $K_{\text{z}}$. As the magnitude of $B$ increases, the magnetization $\overrightarrow{M}$ of both injector and detector FM electrodes makes a finite angle $\theta_i$ with respect to its initial direction (+y-axis), and the injected spins have their quantization axis along $\theta_i$ (Fig.~\ref{osv_steps}(a)). Now, the measured spin-signal $R_{\text{nl}}^{\text{P1}}$ in the parallel configuration can be expressed as:
 \begin{equation}
  R_{\text{nl}}^{\text{P1}} =  R_{\text{nl}}^{||}\cos^2\theta_{\text{i}}\zeta_{||}({B\sin\theta_{\text{B}}}) + R_{\text{nl}}^{\perp}\sin^2\theta_{\text{i}}\zeta_{\perp}({B\cos\theta_{\text{B}}})
  \label{sv pl}
 \end{equation}
Here, $\zeta_{||(\perp)}$ is the functional form for the in-plane (out-of-plane) spin precession of dynamics.

{\bf Step-II:} Due to different widths of the FM electrodes, they have different in-plane anisotropies and different switching fields. At a certain magnetic field, the magnetization of the detector reverses the direction of its y-component. Now, the detector magnetization subtends an angle $\theta_f$ with the negative y-axis (Fig.~\ref{osv_steps}(b)). This activity is seen as a switch due to the direction reversal of both in-plane and out-of-plane magnetization component with respect to its initial orientation.
% and ii) due to the change in the magnitude of both magnetization components because of change in the detector angle magnitude, i.e. $\theta_{\text{i}}\neq \theta_{\text{f}}$ (Fig.~\ref{simulation}(b)). 
The factorization of in-plane and out-of-plane components can be understood via the presented vector diagram in Fig.~\ref{osv_steps}(b) in following steps:
\begin{itemize}
 \item The injector electrode injects the spin signal along $\theta_{\text{i}}$, represented by the blue arrow b-c in Fig.~\ref{osv_steps}(b).  
%  \begin{equation}
%   R_{\text{nl}}^{||}\cos\theta_{\text{i}} + R_{\text{nl}}^{\perp}\sin\theta_{\text{i}},
%  \end{equation}
% if there were no dephasing.  
 \item The detector measures the projection of the injected spin-signal which has its quantization axis at $\theta_{\text{i}}$, along the detector magnetization axis along b-a, shown as a black dashed line in Fig.~\ref{osv_steps}(b). Now the magnetization axis, along which the spin-signal is measured becomes: 
 \begin{equation}
  \overrightarrow{M}_{\text{injector}}^{\text{new}}=-(\cos\theta_{\text{f}}\hat{j} + \sin\theta_{\text{f}}\hat{k})\cos(\theta_{\text{i}}+\theta_{\text{f}}),
 \end{equation}

 where $\hat{j},\hat{k}$ are the unit vectors along y and z-axis, respectively. Since the in-plane and out-of-plane spin-signals have magnitudes $R_{\text{nl}}^{||}\cos\theta_{\text{i}}\zeta_{||}({B\sin\theta_{\text{B}}})$ and $R_{\text{nl}}^{\perp}\sin\theta_{\text{i}}\zeta_{\perp}({B\cos\theta_{\text{B}}})$, the spin-signal measured by the detector becomes:
 
 \begin{multline}
  R_{\text{nl}}^{\text{AP}} = -[R_{\text{nl}}^{||}\cos\theta_{\text{i}}\zeta_{||}({B\sin\theta_{\text{B}}})\cos\theta_{\text{f}} +\\
  R_{\text{nl}}^{\perp}\sin\theta_{\text{i}}\zeta_{\perp}({B\cos\theta_{\text{B}}})\sin\theta_{\text{f}}]\cos(\theta_{\text{i}}+\theta_{\text{f}})
  \label{sv apl}
 \end{multline}
 
\end{itemize}

{\bf Step-III:} Finally, the injector electrode reverses its magnetization and both electrodes have their magnetizations pointing in the same direction, and making an angle $\theta_{\text{f}}$ with the device plane Fig.~\ref{osv_steps}(c). The spin-signal $R_{\text{nl}}^{\text{P2}}$ has the same expression as in Eq.~\ref{sv pl}, except $\theta_{\text{i}}$ is replaced with  $\theta_{\text{f}}$. The desired spin valve signal can be obtained by subtracting Eq.~\ref{sv apl} with Eq.~\ref{sv pl} with appropriate $\theta$ values, obtained from Fig.~\ref{simulation} at corresponding magnetization switching fields. 

A data set for the oblique spin valve measurements is shown in Fig.~\ref{osv}. As expected by the simulation results in Fig.~\ref{simulation}, the magnetization switching follows the relation $B_0\sim B\cos\theta_{\text{B}}$, where $B_0$ is the magnetization switching field $\sim$ 40 mT for the in-plane spin valve (black curve in Fig.~\ref{osv}). The measured signal has contribution from both in-plane and out-of-plane magnetization switching.  As suggested by the simulation results, the magnetic field dependent background in the measurement has similar trend as observed in Fig.~\ref{simulation}(b)) due to the field-dependent magnetization angle, and has contribution of the out-of-plane spin-signal. The processed data after removing this field dependence is shown in Fig.3(a) of the main text which shows a clear enhancement in the measured spin valve signal magnitude. This is a consequence of large spin-life time anisotropy present in the system, and is discussed in the manuscript in detail.

\begin{figure}[!h]
\includegraphics[scale=1]{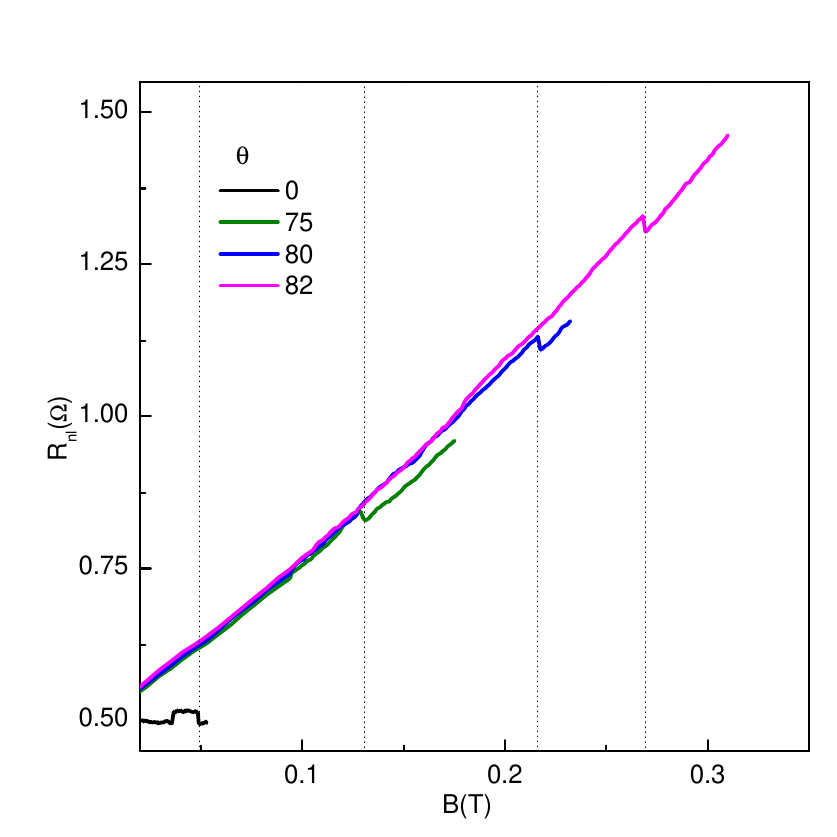}
\caption{\label{osv}(a)  OSV measurements for BLG/WS$_2$ at different $\theta_{\text{B}}$ values for the injector-detector separation $L$=1 $\mu$m with out background removal. Vertical dashed lines indicate the magnitudes of the magnetization switching field magnitudes on x-axis at different $\theta_{\text{B}}$ values. The black curve is the in-plane spin-valve measurement at $\theta_{\text{B}}=0^{\circ}$. The curves measured at $\theta_{\text{B}}=75-82^{\circ}$ have the contribution from both in-plane and out-of-plane spin-signals. The enhanced contribution of the out-of-plane spin-signal component during the magnetization reversal, i.e. enhanced switch magnitude in $R_{\text{nl}}$ for the measurements at higher $\theta_{\text{B}}$ values can be seen clearly in Fig. 4(b) of the main text after the background removal.}
\end{figure}

\begin{figure}[!h]
\includegraphics[scale=1]{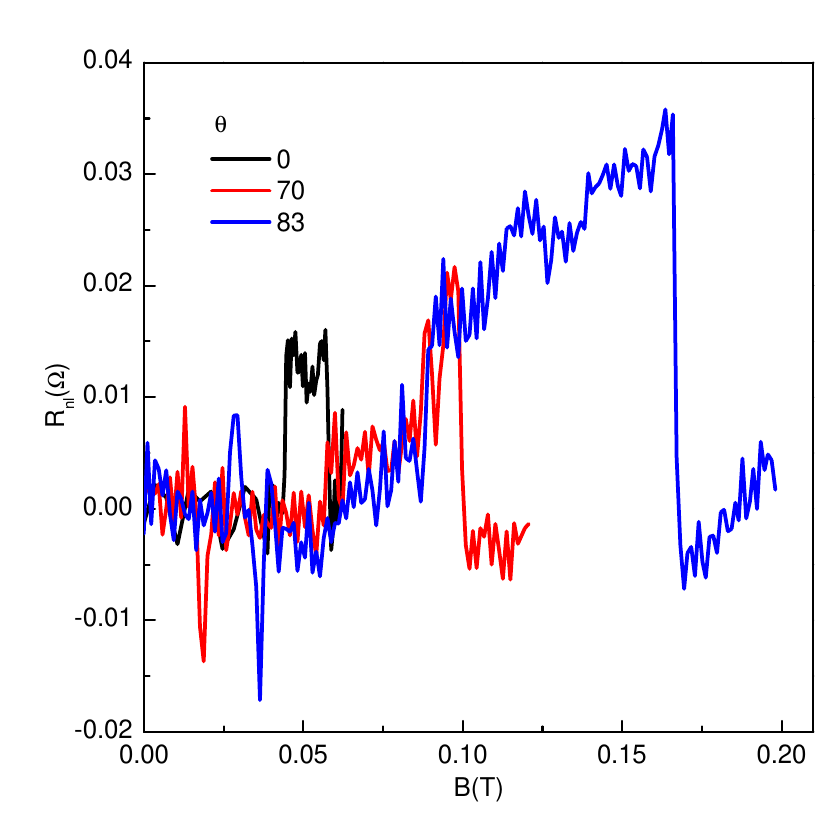}
\caption{\label{osv_processed} Additional OSV measurements at $L$=1 $\mu$m at $V_{\text{bg}} =$ 0 V (stack A). }
\end{figure}

An additional set of OSV measurements for a different region (on the right side) of stack A is shown in Fig.~\ref{osv_processed}. For this set the FM electrodes at $\theta_{\text{B}} =$ 83$^{\circ}$ switch earlier than the expected switching field, i.e. $B_0/\cos\theta_{\text{b}}\sim$ 300 mT, and using the angles obtained in Fig.~\ref{simulation} and $\tau_{||}$ in the region, the analysis yields $\eta\sim$244 and $\tau_{\perp}\sim$ 4 ns.  The overestimation of $\eta$ is probably due to earlier switching of the FM electrode. However, the effect of anisotropy can be clearly seen in the measurement.

\section{Nonlocal Hanle signal versus orbital magnetoresistance}

\begin{figure}[!h]
\includegraphics[scale=1]{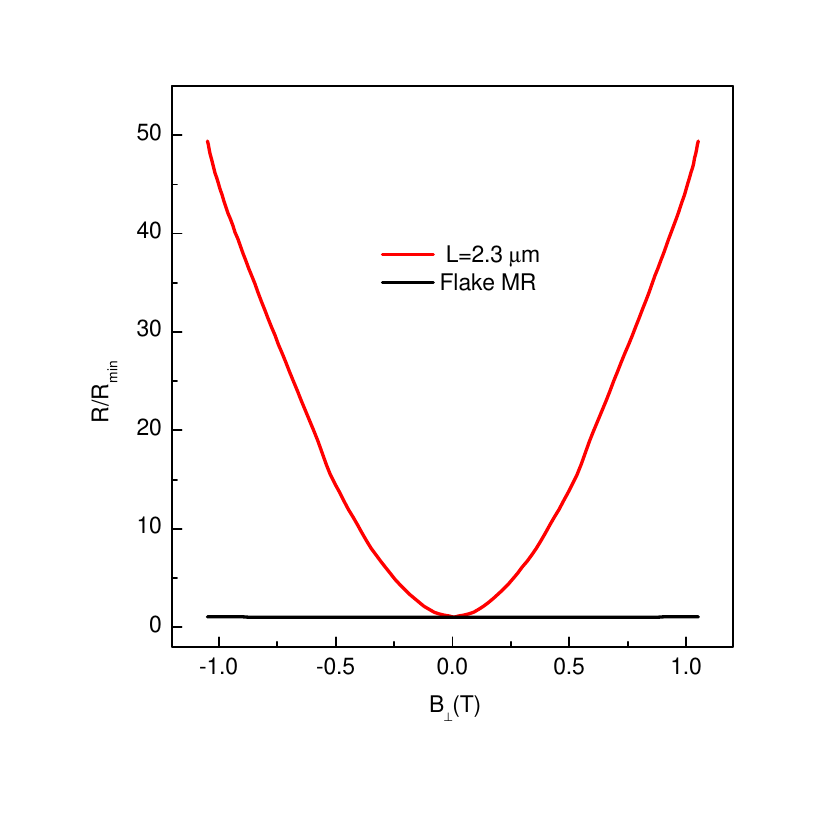}
\caption{\label{Hanle MR} Hanle (parallel configuration) at the injector-detector separation $L$ =2.3 $\mu$m and the flake magnetoresistance (black curve) are symmetrized and normalized with $R_{\text{nl}}^{\text{min}}$ and $R_{\text{MR}}^{\text{min}}$ value in order to emphasize the signal enhancement in the nonlocal configuration.}
\end{figure}

A negligible charge background signal due to the orbital magnetoresistance of the graphene flake is present at the applied $B_{\perp} =$ (Fig.~\ref{Hanle MR}). 
Here, for the same channel $R_{\text{nl}}$ increases almost 50 fold  whereas there is hardly any change in the background MR signal(Fig.~\ref{Hanle MR}). Therefore the observed increase in $R_{\text{nl}}$ at high $B_{\perp}$ is clearly not due to the orbital magnetoresistance of the graphene-flake.
\section{Estimating out-of-plane spin relaxation time via Hanle measurements}
\begin{figure}[!h]
\includegraphics[scale=1]{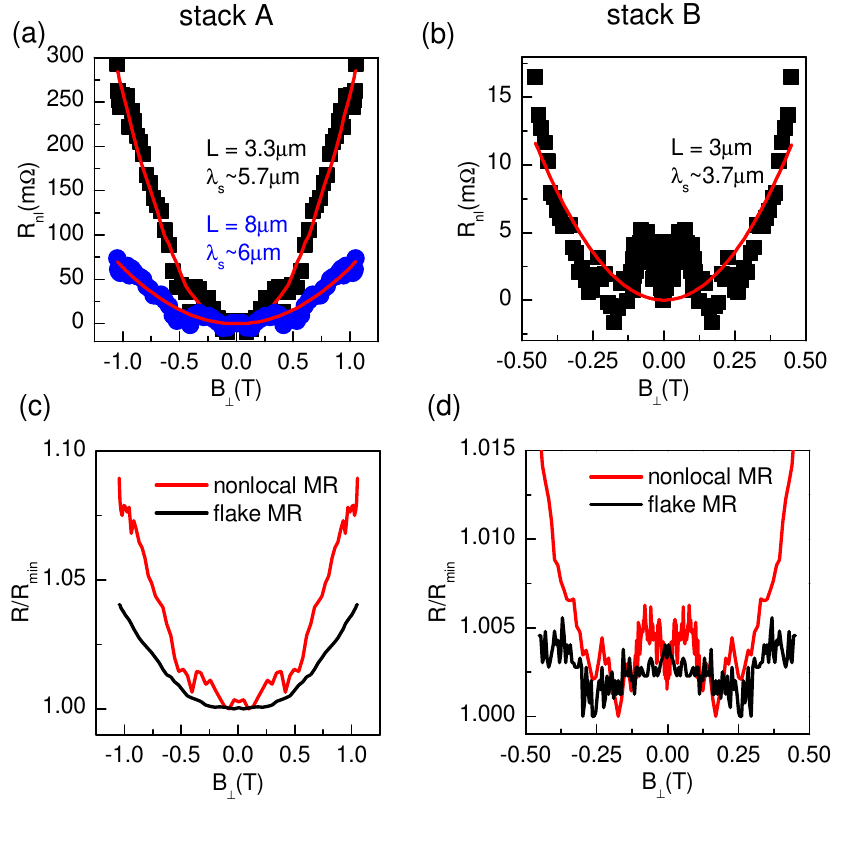}
\caption{\label{fig3} (a) Symmetrized Hanle curves (stack A) and fits (in red) with Eq.~\ref{final total rnl} after subtracting the background-signal of $\sim$3 $\Omega$ and 1 $\Omega$ at two different injector-detector separations, respectively (at $V_{\text{bg}}$ = 0 ) result in similar $\lambda_{\text{s}}^{\perp}$ (b) Additional Hanle measurements and the fit for stack B (after subtracting the background signal $\sim$1 $\Omega$).  Nonlocal resistance and flake magnetoresistance are normalized and plotted together in order to highlight the relative difference between them in (c) for stack A and (d) for stack B.}
\end{figure}
It is already explained in the previous section that at a nonzero magnetic field $B$ applied at an angle $\theta_{\text{B}}$ with the device plane, the magnetization vector $\overrightarrow{M}$ makes a finite angle $\theta$ with the device plane (Fig.~\ref{mgtc angle}). Here, we represent a specific case with $\theta_{\text{B}}=90^{\circ}$ for Hanle measurements. Here, we would represent $B$ as $B_\perp$ and assume that both injector and detector behave identically and their $\overrightarrow{M}$ vectors make same angle $\theta$. At $B_\perp \neq$0, $\overrightarrow{M}$ has its quatization axis not in the device plane, it also electrically injects a nonzero out-of-plane spin-signal. If $\overrightarrow{M}$ for both injector and detector were pointing perpendicular to the device plane, the measured nonlocal signal $R_{\text{nl}}^{\perp}$ would be written as:
\begin{equation}
 R_{\text{nl}}^{\perp}=\frac{P^2 R_{\text{sq}}(B_{\perp})\lambda_{\text{s}}^\perp e^{-\frac{L}{\lambda_{\text{s}}^\perp}}}{2w}.
 \label{only rnl}
\end{equation}

Here $\lambda_{\text{s}}^\perp$ is the spin-relaxation length for the out-of-plane spins in graphene and $P$ is the contact polarization of injector and detector electrodes, which is obtained via in-plane spin-transport measurements. $R_{\text{sq}}(B_{\perp})$ is the magnetoresistance (MR) of the graphene flake in presence of the out-of-plane magnetic field. However, in general $\theta < \pi/2$ for the values of $B_\perp <$ 1.2 T due to limitations of the electromagnet in the setup, we inject and detect only a fraction of $R_{\text{nl}}^{\perp}$ that is proportional to $\sin^2\theta(B_\perp)$, and the in-plane spin-signal $R_{\text{nl}}^{||}$ that is proportional to $\cos^2\theta(B_\perp)$ and gets dephased by $B_{\perp}$.  

FM contacts also measure charge-related MR and a constant spin-independent background due to current spreading and homogeneous current distribution even in the nonlocal part of the circuit. This contribution can be represented as:

\begin{equation}
 R_{\text{nl}}^{\text{ch}} = C1 R_{\text{sq}}(B_{\perp}) + C2
\end{equation}

Therefore, the total measured nonlocal signal $R_{\text{nl}}^{\text{T}}$ is:
\begin{equation}
 R_{\text{nl}}^{\text{T}}(B_\perp) = R_{\text{nl}}^{\perp} \sin^2\theta(B_\perp) \pm R_{\text{nl}}^{||} \cos^2\theta(B_\perp)\times \zeta(B_{\perp}) + R_{\text{nl}}^{\text{ch}}
 \label{contribution}
 \end{equation}
 Here +(-) before the expression for the in-plane spin signal is for P(AP) magnetization configuration of the injector-detector electrodes and $\zeta(B_{\perp})$ is the expression for Hanle precession dynamics. The second term can be omitted from Eq.~\ref{contribution} by measuring $R_{\text{nl}}^{\text{T}}(B_\perp)$ for both P and AP configurations of FM electrodes and then averaging them out. Via this exercise, we get rid of the in-plane spin signal and get the following expression:
 \begin{equation}
  R_{\text{nl}}^{\text{T}}(B_\perp) = R_{\text{nl}}^{\perp} \sin^2\theta(B_\perp) + C1 R_{\text{sq}}(B_{\perp}) + C2
  \label{total rnl}
\end{equation}

At $B_\perp$ = 0 T, $R_{\text{sq}}(B_\perp=0)=R_{\text{sq}}$ and $\theta(B_\perp = 0) = 0$, Eq.~\ref{total rnl} reduces to:

\begin{equation}
 R_{\text{nl}}^{\text{T}}(0) = C1 R_{\text{sq}} + C2 
 \label{rnl bg}
\end{equation}
 By subtracting Eq.~\ref{rnl bg} to Eq.~\ref{total rnl} and dividing the resulting expression with $R_{\text{sq}}(B_\perp)$, we obtain:
 
 \begin{equation}
  \frac{R_{\text{nl}}^{\text{T}}(B_\perp)-R_{\text{nl}}^{\text{T}}(0)}{R_{\text{sq}}(B_\perp)} = \frac{R_{\text{nl}}^{\perp} \sin^2\theta(B_\perp)}{R_{\text{sq}}(B_\perp)} + C1\frac{R_{\text{sq}}(B_{\perp})-R_{\text{sq}}}{R_{\text{sq}}(B_\perp)}
  \label{mod total rnl}
 \end{equation}
 Using Eq.~\ref{only rnl}, we obtain the final expression: 

 \begin{equation}
  \frac{R_{\text{nl}}^{\text{T}}(B_\perp)-R_{\text{nl}}^{\text{T}}(0)}{R_{\text{sq}}(B_\perp)} = \frac{p^2 \lambda_{\text{s}}^\perp e^{-\frac{L}{\lambda_{\text{s}}^\perp}}\sin^2\theta(B_\perp)}{2w} + C1\frac{R_{\text{sq}}(B_{\perp})-R_{\text{sq}}}{R_{\text{sq}}(B_\perp)},
  \label{final total rnl}
 \end{equation}

and use it for extracting $\lambda_{\text{s}}^\perp$ and the constant $C1$ which is the fraction of flake MR contributing to the nonlocal signal. Here, $\theta$ is obtained via simulations, following the procedure mentioned earlier using $\theta_{\text{B}}=\pi/2$.  Experimental data of $R_{\text{nl}}$ and the fit with Eq.~\ref{final total rnl} is shown in Fig.~\ref{fig3}(a,b).   
% \begin{equation}
%  R_{nl}(B^\perp)=R_{nl}(B^\perp=0)+C\times R_{\text{sq}}(B^\perp)+R_{\text{nl}^{||}}(B^\perp=0)\zeta(B^\perp)cos^2(\theta)+\frac{p^2\times R_{\text{sq}}((B^\perp))\times \lambda_{\text{s}}^\perp\times e^{-\frac{L}{\lambda{\text{s}}}}}{2w}
% \end{equation}
% \section{Additional OSV measurements}
\section{Estimation of Valley-Zeeman and Rashba SOC strengths}

In graphene/TMD heterostructures, different spin-orbit coupling strengths are induced in graphene in the in-plane and out-of-plane directions because of weak van der Waals interactions with the contacting TMD \cite{cummings_giant_2017}. This effect can be measured in the anisotropy of in-plane ($\tau_{||}$) and out-of-plane spin-relaxation time ($\tau_{\perp}$) using the following relation:
\begin{equation}
 \eta=\frac{\tau_{\perp}}{\tau_{||}} \sim \bigg(\frac{\lambda_{VZ}}{\lambda_{R}}\bigg)^2\frac{\tau_{\text{iv}}}{\tau_{\text{p}}}.
\end{equation}

Here $\lambda_{VZ}$ and $\lambda_{R}$ are spin-orbit coupling strengths corresponding to the out-of-plane and in-plane spin-orbit field, respectively. $\tau_{\text{iv}}$ is the intervalley scattering time, and $\tau_{\text{p}}$ is the momentum relaxation time of electron.

From the charge and spin transport measurements, we obtain the diffusion coefficient $D\sim$0.01-0.03 m$^2$V$^{-1}$s$^{-1}$. Following the relation $D\sim v_{\text{F}}^2\tau_{\text{p}}$, where $v_{\text{F}}$=10$^6$ m/s is the Fermi velocity of electrons in graphene, we obtain $\tau_{\text{p}}\sim$ 0.01-0.03 ps. Typically, for strong intervalley scattering, we can assume the relation $\tau_{\text{iv}}\sim5\tau_{\text{p}}$ \cite{cummings_giant_2017}, and estimate $\tau_{\text{iv}}\sim$ 0.05-0.15 ps. From the spin-transport experiments, we already know $\tau_{\perp} \sim$ 1 ns and $\tau_{||} \sim$ 30 ps. We can now estimate $\lambda_R$ and $\lambda_{VZ}$ independently by assuming that the spin-relaxation is dominated by the Dyakonov Perel mechanism \cite{cummings_giant_2017}, i.e. using the relations $\tau_{\perp}^{-1}=(2\lambda_{R}/\hbar)^2\tau_{\text{p}}$ and $\tau_{||}^{-1}=(2\lambda_{VZ}/\hbar)^2\tau_{\text{iv}}$, respectively. We obtain $\lambda_{\text{R}} \sim$ 100$\mu eV$ and  $\lambda_{\text{VZ}} \sim$ 350$\mu eV$. The obtain values are of similar order magnitude as reported in literature \cite{omar_spin_2018, zihlmann_large_2018, cummings_giant_2017}. 

% \bibliography{supplementary}

\begin{thebibliography}{44}%
\makeatletter
\providecommand \@ifxundefined [1]{%
 \@ifx{#1\undefined}
}%
\providecommand \@ifnum [1]{%
 \ifnum #1\expandafter \@firstoftwo
 \else \expandafter \@secondoftwo
 \fi
}%
\providecommand \@ifx [1]{%
 \ifx #1\expandafter \@firstoftwo
 \else \expandafter \@secondoftwo
 \fi
}%
\providecommand \natexlab [1]{#1}%
\providecommand \enquote  [1]{``#1''}%
\providecommand \bibnamefont  [1]{#1}%
\providecommand \bibfnamefont [1]{#1}%
\providecommand \citenamefont [1]{#1}%
\providecommand \href@noop [0]{\@secondoftwo}%
\providecommand \href [0]{\begingroup \@sanitize@url \@href}%
\providecommand \@href[1]{\@@startlink{#1}\@@href}%
\providecommand \@@href[1]{\endgroup#1\@@endlink}%
\providecommand \@sanitize@url [0]{\catcode `\\12\catcode `\$12\catcode
  `\&12\catcode `\#12\catcode `\^12\catcode `\_12\catcode `\%12\relax}%
\providecommand \@@startlink[1]{}%
\providecommand \@@endlink[0]{}%
\providecommand \url  [0]{\begingroup\@sanitize@url \@url }%
\providecommand \@url [1]{\endgroup\@href {#1}{\urlprefix }}%
\providecommand \urlprefix  [0]{URL }%
\providecommand \Eprint [0]{\href }%
\providecommand \doibase [0]{http://dx.doi.org/}%
\providecommand \selectlanguage [0]{\@gobble}%
\providecommand \bibinfo  [0]{\@secondoftwo}%
\providecommand \bibfield  [0]{\@secondoftwo}%
\providecommand \translation [1]{[#1]}%
\providecommand \BibitemOpen [0]{}%
\providecommand \bibitemStop [0]{}%
\providecommand \bibitemNoStop [0]{.\EOS\space}%
\providecommand \EOS [0]{\spacefactor3000\relax}%
\providecommand \BibitemShut  [1]{\csname bibitem#1\endcsname}%
\let\auto@bib@innerbib\@empty
%</preamble>
\bibitem [{\citenamefont {Wang}\ \emph {et~al.}(2016)\citenamefont {Wang},
  \citenamefont {Ki}, \citenamefont {Khoo}, \citenamefont {Mauro},
  \citenamefont {Berger}, \citenamefont {Levitov},\ and\ \citenamefont
  {Morpurgo}}]{wang_origin_2016}%
  \BibitemOpen
  \bibfield  {author} {\bibinfo {author} {\bibfnamefont {Z.}~\bibnamefont
  {Wang}}, \bibinfo {author} {\bibfnamefont {D.-K.}\ \bibnamefont {Ki}},
  \bibinfo {author} {\bibfnamefont {J.~Y.}\ \bibnamefont {Khoo}}, \bibinfo
  {author} {\bibfnamefont {D.}~\bibnamefont {Mauro}}, \bibinfo {author}
  {\bibfnamefont {H.}~\bibnamefont {Berger}}, \bibinfo {author} {\bibfnamefont
  {L.~S.}\ \bibnamefont {Levitov}}, \ and\ \bibinfo {author} {\bibfnamefont
  {A.~F.}\ \bibnamefont {Morpurgo}},\ }\href {\doibase
  10.1103/PhysRevX.6.041020} {\bibfield  {journal} {\bibinfo  {journal} {Phys.
  Rev. X}\ }\textbf {\bibinfo {volume} {6}},\ \bibinfo {pages} {041020}
  (\bibinfo {year} {2016})}\BibitemShut {NoStop}%
\bibitem [{\citenamefont {Gmitra}\ and\ \citenamefont
  {Fabian}(2015)}]{gmitra_graphene_2015}%
  \BibitemOpen
  \bibfield  {author} {\bibinfo {author} {\bibfnamefont {M.}~\bibnamefont
  {Gmitra}}\ and\ \bibinfo {author} {\bibfnamefont {J.}~\bibnamefont
  {Fabian}},\ }\href {\doibase 10.1103/PhysRevB.92.155403} {\bibfield
  {journal} {\bibinfo  {journal} {Phys. Rev. B}\ }\textbf {\bibinfo {volume}
  {92}},\ \bibinfo {pages} {155403} (\bibinfo {year} {2015})}\BibitemShut
  {NoStop}%
\bibitem [{\citenamefont {Omar}\ and\ \citenamefont {van
  Wees}(2018)}]{omar_spin_2018}%
  \BibitemOpen
  \bibfield  {author} {\bibinfo {author} {\bibfnamefont {S.}~\bibnamefont
  {Omar}}\ and\ \bibinfo {author} {\bibfnamefont {B.~J.}\ \bibnamefont {van
  Wees}},\ }\href {\doibase 10.1103/PhysRevB.97.045414} {\bibfield  {journal}
  {\bibinfo  {journal} {Phys. Rev. B}\ }\textbf {\bibinfo {volume} {97}},\
  \bibinfo {pages} {045414} (\bibinfo {year} {2018})}\BibitemShut {NoStop}%
\bibitem [{\citenamefont {Xiao}\ \emph {et~al.}(2007)\citenamefont {Xiao},
  \citenamefont {Yao},\ and\ \citenamefont
  {Niu}}]{xiao_valley-contrasting_2007}%
  \BibitemOpen
  \bibfield  {author} {\bibinfo {author} {\bibfnamefont {D.}~\bibnamefont
  {Xiao}}, \bibinfo {author} {\bibfnamefont {W.}~\bibnamefont {Yao}}, \ and\
  \bibinfo {author} {\bibfnamefont {Q.}~\bibnamefont {Niu}},\ }\href {\doibase
  10.1103/PhysRevLett.99.236809} {\bibfield  {journal} {\bibinfo  {journal}
  {Phys. Rev. Lett.}\ }\textbf {\bibinfo {volume} {99}},\ \bibinfo {pages}
  {236809} (\bibinfo {year} {2007})}\BibitemShut {NoStop}%
\bibitem [{\citenamefont {Leutenantsmeyer}\ \emph {et~al.}(2018)\citenamefont
  {Leutenantsmeyer}, \citenamefont {Ingla-Ayn\'{e}s}, \citenamefont {Fabian},\
  and\ \citenamefont {van Wees}}]{leutenantsmeyer_observation_2018}%
  \BibitemOpen
  \bibfield  {author} {\bibinfo {author} {\bibfnamefont {J.~C.}\ \bibnamefont
  {Leutenantsmeyer}}, \bibinfo {author} {\bibfnamefont {J.}~\bibnamefont
  {Ingla-Ayn\'{e}s}}, \bibinfo {author} {\bibfnamefont {J.}~\bibnamefont
  {Fabian}}, \ and\ \bibinfo {author} {\bibfnamefont {B.~J.}\ \bibnamefont {van
  Wees}},\ }\href {\doibase 10.1103/PhysRevLett.121.127702} {\bibfield
  {journal} {\bibinfo  {journal} {Phys. Rev. Lett.}\ }\textbf {\bibinfo
  {volume} {121}},\ \bibinfo {pages} {127702} (\bibinfo {year}
  {2018})}\BibitemShut {NoStop}%
\bibitem [{\citenamefont {Xu}\ \emph {et~al.}(2018)\citenamefont {Xu},
  \citenamefont {Zhu}, \citenamefont {Luo}, \citenamefont {Lu},\ and\
  \citenamefont {Kawakami}}]{xu_strong_2018}%
  \BibitemOpen
  \bibfield  {author} {\bibinfo {author} {\bibfnamefont {J.}~\bibnamefont
  {Xu}}, \bibinfo {author} {\bibfnamefont {T.}~\bibnamefont {Zhu}}, \bibinfo
  {author} {\bibfnamefont {Y.~K.}\ \bibnamefont {Luo}}, \bibinfo {author}
  {\bibfnamefont {Y.-M.}\ \bibnamefont {Lu}}, \ and\ \bibinfo {author}
  {\bibfnamefont {R.~K.}\ \bibnamefont {Kawakami}},\ }\href {\doibase
  10.1103/PhysRevLett.121.127703} {\bibfield  {journal} {\bibinfo  {journal}
  {Phys. Rev. Lett.}\ }\textbf {\bibinfo {volume} {121}},\ \bibinfo {pages}
  {127703} (\bibinfo {year} {2018})}\BibitemShut {NoStop}%
\bibitem [{\citenamefont {Zihlmann}\ \emph {et~al.}(2018)\citenamefont
  {Zihlmann}, \citenamefont {Cummings}, \citenamefont {Garcia}, \citenamefont
  {Kedves}, \citenamefont {Watanabe}, \citenamefont {Taniguchi}, \citenamefont
  {Sch\"{o}nenberger},\ and\ \citenamefont {Makk}}]{zihlmann_large_2018}%
  \BibitemOpen
  \bibfield  {author} {\bibinfo {author} {\bibfnamefont {S.}~\bibnamefont
  {Zihlmann}}, \bibinfo {author} {\bibfnamefont {A.~W.}\ \bibnamefont
  {Cummings}}, \bibinfo {author} {\bibfnamefont {J.~H.}\ \bibnamefont
  {Garcia}}, \bibinfo {author} {\bibfnamefont {M.}~\bibnamefont {Kedves}},
  \bibinfo {author} {\bibfnamefont {K.}~\bibnamefont {Watanabe}}, \bibinfo
  {author} {\bibfnamefont {T.}~\bibnamefont {Taniguchi}}, \bibinfo {author}
  {\bibfnamefont {C.}~\bibnamefont {Sch\"{o}nenberger}}, \ and\ \bibinfo
  {author} {\bibfnamefont {P.}~\bibnamefont {Makk}},\ }\href {\doibase
  10.1103/PhysRevB.97.075434} {\bibfield  {journal} {\bibinfo  {journal} {Phys.
  Rev. B}\ }\textbf {\bibinfo {volume} {97}},\ \bibinfo {pages} {075434}
  (\bibinfo {year} {2018})}\BibitemShut {NoStop}%
\bibitem [{\citenamefont {Cummings}\ \emph {et~al.}(2017)\citenamefont
  {Cummings}, \citenamefont {Garcia}, \citenamefont {Fabian},\ and\
  \citenamefont {Roche}}]{cummings_giant_2017}%
  \BibitemOpen
  \bibfield  {author} {\bibinfo {author} {\bibfnamefont {A.~W.}\ \bibnamefont
  {Cummings}}, \bibinfo {author} {\bibfnamefont {J.~H.}\ \bibnamefont
  {Garcia}}, \bibinfo {author} {\bibfnamefont {J.}~\bibnamefont {Fabian}}, \
  and\ \bibinfo {author} {\bibfnamefont {S.}~\bibnamefont {Roche}},\ }\href
  {\doibase 10.1103/PhysRevLett.119.206601} {\bibfield  {journal} {\bibinfo
  {journal} {Phys. Rev. Lett.}\ }\textbf {\bibinfo {volume} {119}},\ \bibinfo
  {pages} {206601} (\bibinfo {year} {2017})}\BibitemShut {NoStop}%
\bibitem [{\citenamefont {Ghiasi}\ \emph {et~al.}(2017)\citenamefont {Ghiasi},
  \citenamefont {Ingla-Ayn\'{e}s}, \citenamefont {Kaverzin},\ and\
  \citenamefont {van Wees}}]{ghiasi_large_2017}%
  \BibitemOpen
  \bibfield  {author} {\bibinfo {author} {\bibfnamefont {T.~S.}\ \bibnamefont
  {Ghiasi}}, \bibinfo {author} {\bibfnamefont {J.}~\bibnamefont
  {Ingla-Ayn\'{e}s}}, \bibinfo {author} {\bibfnamefont {A.~A.}\ \bibnamefont
  {Kaverzin}}, \ and\ \bibinfo {author} {\bibfnamefont {B.~J.}\ \bibnamefont
  {van Wees}},\ }\href
  {https://pubs.acs.org/doi/abs/10.1021/acs.nanolett.7b03460} {\bibfield
  {journal} {\bibinfo  {journal} {Nano Lett.}\ }\textbf {\bibinfo {volume}
  {17}} (\bibinfo {year} {2017})}\BibitemShut {NoStop}%
\bibitem [{\citenamefont {Safeer}\ \emph {et~al.}(2019)\citenamefont {Safeer},
  \citenamefont {Ingla-Ayn\'{e}s}, \citenamefont {Herling}, \citenamefont
  {Garcia}, \citenamefont {Vila}, \citenamefont {Ontoso}, \citenamefont
  {Calvo}, \citenamefont {Roche}, \citenamefont {Hueso},\ and\ \citenamefont
  {Casanova}}]{safeer_room-temperature_2019}%
  \BibitemOpen
  \bibfield  {author} {\bibinfo {author} {\bibfnamefont {C.~K.}\ \bibnamefont
  {Safeer}}, \bibinfo {author} {\bibfnamefont {J.}~\bibnamefont
  {Ingla-Ayn\'{e}s}}, \bibinfo {author} {\bibfnamefont {F.}~\bibnamefont
  {Herling}}, \bibinfo {author} {\bibfnamefont {J.~H.}\ \bibnamefont {Garcia}},
  \bibinfo {author} {\bibfnamefont {M.}~\bibnamefont {Vila}}, \bibinfo {author}
  {\bibfnamefont {N.}~\bibnamefont {Ontoso}}, \bibinfo {author} {\bibfnamefont
  {M.~R.}\ \bibnamefont {Calvo}}, \bibinfo {author} {\bibfnamefont
  {S.}~\bibnamefont {Roche}}, \bibinfo {author} {\bibfnamefont {L.~E.}\
  \bibnamefont {Hueso}}, \ and\ \bibinfo {author} {\bibfnamefont
  {F.}~\bibnamefont {Casanova}},\ }\href {\doibase
  10.1021/acs.nanolett.8b04368} {\bibfield  {journal} {\bibinfo  {journal}
  {Nano Lett.}\ }\textbf {\bibinfo {volume} {19}},\ \bibinfo {pages} {1074}
  (\bibinfo {year} {2019})}\BibitemShut {NoStop}%
\bibitem [{\citenamefont {Garcia}\ \emph {et~al.}(2017)\citenamefont {Garcia},
  \citenamefont {Cummings},\ and\ \citenamefont {Roche}}]{garcia_spin_2017}%
  \BibitemOpen
  \bibfield  {author} {\bibinfo {author} {\bibfnamefont {J.~H.}\ \bibnamefont
  {Garcia}}, \bibinfo {author} {\bibfnamefont {A.~W.}\ \bibnamefont
  {Cummings}}, \ and\ \bibinfo {author} {\bibfnamefont {S.}~\bibnamefont
  {Roche}},\ }\href {\doibase 10.1021/acs.nanolett.7b02364} {\bibfield
  {journal} {\bibinfo  {journal} {Nano Lett.}\ }\textbf {\bibinfo {volume}
  {17}},\ \bibinfo {pages} {5078} (\bibinfo {year} {2017})}\BibitemShut
  {NoStop}%
\bibitem [{\citenamefont {Song}\ \emph {et~al.}(2017)\citenamefont {Song},
  \citenamefont {Zhang}, \citenamefont {Su}, \citenamefont {Yuan},
  \citenamefont {Chen}, \citenamefont {Xing}, \citenamefont {Shi},
  \citenamefont {Sun},\ and\ \citenamefont {Han}}]{song_observation_2017}%
  \BibitemOpen
  \bibfield  {author} {\bibinfo {author} {\bibfnamefont {Q.}~\bibnamefont
  {Song}}, \bibinfo {author} {\bibfnamefont {H.}~\bibnamefont {Zhang}},
  \bibinfo {author} {\bibfnamefont {T.}~\bibnamefont {Su}}, \bibinfo {author}
  {\bibfnamefont {W.}~\bibnamefont {Yuan}}, \bibinfo {author} {\bibfnamefont
  {Y.}~\bibnamefont {Chen}}, \bibinfo {author} {\bibfnamefont {W.}~\bibnamefont
  {Xing}}, \bibinfo {author} {\bibfnamefont {J.}~\bibnamefont {Shi}}, \bibinfo
  {author} {\bibfnamefont {J.}~\bibnamefont {Sun}}, \ and\ \bibinfo {author}
  {\bibfnamefont {W.}~\bibnamefont {Han}},\ }\href {\doibase
  10.1126/sciadv.1602312} {\bibfield  {journal} {\bibinfo  {journal} {Sci.
  Adv.}\ }\textbf {\bibinfo {volume} {3}},\ \bibinfo {pages} {1602312}
  (\bibinfo {year} {2017})}\BibitemShut {NoStop}%
\bibitem [{\citenamefont {Soumyanarayanan}\ \emph {et~al.}(2016)\citenamefont
  {Soumyanarayanan}, \citenamefont {Reyren}, \citenamefont {Fert},\ and\
  \citenamefont {Panagopoulos}}]{soumyanarayanan_emergent_2016}%
  \BibitemOpen
  \bibfield  {author} {\bibinfo {author} {\bibfnamefont {A.}~\bibnamefont
  {Soumyanarayanan}}, \bibinfo {author} {\bibfnamefont {N.}~\bibnamefont
  {Reyren}}, \bibinfo {author} {\bibfnamefont {A.}~\bibnamefont {Fert}}, \ and\
  \bibinfo {author} {\bibfnamefont {C.}~\bibnamefont {Panagopoulos}},\ }\href
  {\doibase 10.1038/nature19820} {\bibfield  {journal} {\bibinfo  {journal}
  {Nature}\ }\textbf {\bibinfo {volume} {539}},\ \bibinfo {pages} {509}
  (\bibinfo {year} {2016})}\BibitemShut {NoStop}%
\bibitem [{\citenamefont {Isasa}\ \emph {et~al.}(2016)\citenamefont {Isasa},
  \citenamefont {Mart\'{i}nez-Velarte}, \citenamefont {Villamor}, \citenamefont
  {Mag\'{e}n}, \citenamefont {Morell\'{o}n}, \citenamefont {De~Teresa},
  \citenamefont {Ibarra}, \citenamefont {Vignale}, \citenamefont {Chulkov},
  \citenamefont {Krasovskii}, \citenamefont {Hueso},\ and\ \citenamefont
  {Casanova}}]{isasa_origin_2016}%
  \BibitemOpen
  \bibfield  {author} {\bibinfo {author} {\bibfnamefont {M.}~\bibnamefont
  {Isasa}}, \bibinfo {author} {\bibfnamefont {M.~C.}\ \bibnamefont
  {Mart\'{i}nez-Velarte}}, \bibinfo {author} {\bibfnamefont {E.}~\bibnamefont
  {Villamor}}, \bibinfo {author} {\bibfnamefont {C.}~\bibnamefont {Mag\'{e}n}},
  \bibinfo {author} {\bibfnamefont {L.}~\bibnamefont {Morell\'{o}n}}, \bibinfo
  {author} {\bibfnamefont {J.~M.}\ \bibnamefont {De~Teresa}}, \bibinfo {author}
  {\bibfnamefont {M.~R.}\ \bibnamefont {Ibarra}}, \bibinfo {author}
  {\bibfnamefont {G.}~\bibnamefont {Vignale}}, \bibinfo {author} {\bibfnamefont
  {E.~V.}\ \bibnamefont {Chulkov}}, \bibinfo {author} {\bibfnamefont {E.~E.}\
  \bibnamefont {Krasovskii}}, \bibinfo {author} {\bibfnamefont {L.~E.}\
  \bibnamefont {Hueso}}, \ and\ \bibinfo {author} {\bibfnamefont
  {F.}~\bibnamefont {Casanova}},\ }\href {\doibase 10.1103/PhysRevB.93.014420}
  {\bibfield  {journal} {\bibinfo  {journal} {Phys. Rev. B}\ }\textbf {\bibinfo
  {volume} {93}},\ \bibinfo {pages} {014420} (\bibinfo {year}
  {2016})}\BibitemShut {NoStop}%
\bibitem [{\citenamefont {S\'{a}nchez}\ \emph {et~al.}(2013)\citenamefont
  {S\'{a}nchez}, \citenamefont {Vila}, \citenamefont {Desfonds}, \citenamefont
  {Gambarelli}, \citenamefont {Attan\'{e}}, \citenamefont {De~Teresa},
  \citenamefont {Mag\'{e}n},\ and\ \citenamefont
  {Fert}}]{sanchez_spin--charge_2013}%
  \BibitemOpen
  \bibfield  {author} {\bibinfo {author} {\bibfnamefont {J.~C.~R.}\
  \bibnamefont {S\'{a}nchez}}, \bibinfo {author} {\bibfnamefont
  {L.}~\bibnamefont {Vila}}, \bibinfo {author} {\bibfnamefont {G.}~\bibnamefont
  {Desfonds}}, \bibinfo {author} {\bibfnamefont {S.}~\bibnamefont
  {Gambarelli}}, \bibinfo {author} {\bibfnamefont {J.~P.}\ \bibnamefont
  {Attan\'{e}}}, \bibinfo {author} {\bibfnamefont {J.~M.}\ \bibnamefont
  {De~Teresa}}, \bibinfo {author} {\bibfnamefont {C.}~\bibnamefont
  {Mag\'{e}n}}, \ and\ \bibinfo {author} {\bibfnamefont {A.}~\bibnamefont
  {Fert}},\ }\href {\doibase 10.1038/ncomms3944} {\bibfield  {journal}
  {\bibinfo  {journal} {Nat. Commun.}\ }\textbf {\bibinfo {volume} {4}},\
  \bibinfo {pages} {2944} (\bibinfo {year} {2013})}\BibitemShut {NoStop}%
\bibitem [{\citenamefont {Shen}\ \emph {et~al.}(2014)\citenamefont {Shen},
  \citenamefont {Vignale},\ and\ \citenamefont
  {Raimondi}}]{shen_microscopic_2014}%
  \BibitemOpen
  \bibfield  {author} {\bibinfo {author} {\bibfnamefont {K.}~\bibnamefont
  {Shen}}, \bibinfo {author} {\bibfnamefont {G.}~\bibnamefont {Vignale}}, \
  and\ \bibinfo {author} {\bibfnamefont {R.}~\bibnamefont {Raimondi}},\ }\href
  {\doibase 10.1103/PhysRevLett.112.096601} {\bibfield  {journal} {\bibinfo
  {journal} {Phys. Rev. Lett.}\ }\textbf {\bibinfo {volume} {112}},\ \bibinfo
  {pages} {096601} (\bibinfo {year} {2014})}\BibitemShut {NoStop}%
\bibitem [{\citenamefont {Kane}\ and\ \citenamefont
  {Mele}(2005)}]{kane_$z_2$_2005}%
  \BibitemOpen
  \bibfield  {author} {\bibinfo {author} {\bibfnamefont {C.~L.}\ \bibnamefont
  {Kane}}\ and\ \bibinfo {author} {\bibfnamefont {E.~J.}\ \bibnamefont
  {Mele}},\ }\href {\doibase 10.1103/PhysRevLett.95.146802} {\bibfield
  {journal} {\bibinfo  {journal} {Phys. Rev. Lett.}\ }\textbf {\bibinfo
  {volume} {95}},\ \bibinfo {pages} {146802} (\bibinfo {year}
  {2005})}\BibitemShut {NoStop}%
\bibitem [{\citenamefont {Yang}\ \emph {et~al.}(2016)\citenamefont {Yang},
  \citenamefont {Tu}, \citenamefont {Kim}, \citenamefont {Wu}, \citenamefont
  {Wang}, \citenamefont {Alicea}, \citenamefont {Wu}, \citenamefont
  {Bockrath},\ and\ \citenamefont {Shi}}]{yang_tunable_2016}%
  \BibitemOpen
  \bibfield  {author} {\bibinfo {author} {\bibfnamefont {B.}~\bibnamefont
  {Yang}}, \bibinfo {author} {\bibfnamefont {M.-F.}\ \bibnamefont {Tu}},
  \bibinfo {author} {\bibfnamefont {J.}~\bibnamefont {Kim}}, \bibinfo {author}
  {\bibfnamefont {Y.}~\bibnamefont {Wu}}, \bibinfo {author} {\bibfnamefont
  {H.}~\bibnamefont {Wang}}, \bibinfo {author} {\bibfnamefont {J.}~\bibnamefont
  {Alicea}}, \bibinfo {author} {\bibfnamefont {R.}~\bibnamefont {Wu}}, \bibinfo
  {author} {\bibfnamefont {M.}~\bibnamefont {Bockrath}}, \ and\ \bibinfo
  {author} {\bibfnamefont {J.}~\bibnamefont {Shi}},\ }\href {\doibase
  10.1088/2053-1583/3/3/031012} {\bibfield  {journal} {\bibinfo  {journal} {2D
  Mater.}\ }\textbf {\bibinfo {volume} {3}},\ \bibinfo {pages} {031012}
  (\bibinfo {year} {2016})}\BibitemShut {NoStop}%
\bibitem [{\citenamefont {Frank}\ \emph {et~al.}(2018)\citenamefont {Frank},
  \citenamefont {H\"{o}gl}, \citenamefont {Gmitra}, \citenamefont {Kochan},\
  and\ \citenamefont {Fabian}}]{frank_protected_2018}%
  \BibitemOpen
  \bibfield  {author} {\bibinfo {author} {\bibfnamefont {T.}~\bibnamefont
  {Frank}}, \bibinfo {author} {\bibfnamefont {P.}~\bibnamefont {H\"{o}gl}},
  \bibinfo {author} {\bibfnamefont {M.}~\bibnamefont {Gmitra}}, \bibinfo
  {author} {\bibfnamefont {D.}~\bibnamefont {Kochan}}, \ and\ \bibinfo {author}
  {\bibfnamefont {J.}~\bibnamefont {Fabian}},\ }\href {\doibase
  10.1103/PhysRevLett.120.156402} {\bibfield  {journal} {\bibinfo  {journal}
  {Phys. Rev. Lett.}\ }\textbf {\bibinfo {volume} {120}},\ \bibinfo {pages}
  {156402} (\bibinfo {year} {2018})}\BibitemShut {NoStop}%
\bibitem [{\citenamefont {Du}\ \emph {et~al.}(2018)\citenamefont {Du},
  \citenamefont {Zhang}, \citenamefont {Gong}, \citenamefont {Liao},
  \citenamefont {Zhu}, \citenamefont {Yu}, \citenamefont {He}, \citenamefont
  {Liu}, \citenamefont {Yang}, \citenamefont {Shi}, \citenamefont {Gu},
  \citenamefont {Yan}, \citenamefont {Zhang},\ and\ \citenamefont
  {Zhang}}]{du_robust_2018}%
  \BibitemOpen
  \bibfield  {author} {\bibinfo {author} {\bibfnamefont {L.}~\bibnamefont
  {Du}}, \bibinfo {author} {\bibfnamefont {Q.}~\bibnamefont {Zhang}}, \bibinfo
  {author} {\bibfnamefont {B.}~\bibnamefont {Gong}}, \bibinfo {author}
  {\bibfnamefont {M.}~\bibnamefont {Liao}}, \bibinfo {author} {\bibfnamefont
  {J.}~\bibnamefont {Zhu}}, \bibinfo {author} {\bibfnamefont {H.}~\bibnamefont
  {Yu}}, \bibinfo {author} {\bibfnamefont {R.}~\bibnamefont {He}}, \bibinfo
  {author} {\bibfnamefont {K.}~\bibnamefont {Liu}}, \bibinfo {author}
  {\bibfnamefont {R.}~\bibnamefont {Yang}}, \bibinfo {author} {\bibfnamefont
  {D.}~\bibnamefont {Shi}}, \bibinfo {author} {\bibfnamefont {L.}~\bibnamefont
  {Gu}}, \bibinfo {author} {\bibfnamefont {F.}~\bibnamefont {Yan}}, \bibinfo
  {author} {\bibfnamefont {G.}~\bibnamefont {Zhang}}, \ and\ \bibinfo {author}
  {\bibfnamefont {Q.}~\bibnamefont {Zhang}},\ }\href {\doibase
  10.1103/PhysRevB.97.115445} {\bibfield  {journal} {\bibinfo  {journal} {Phys.
  Rev. B}\ }\textbf {\bibinfo {volume} {97}},\ \bibinfo {pages} {115445}
  (\bibinfo {year} {2018})}\BibitemShut {NoStop}%
\bibitem [{\citenamefont {Island}\ \emph {et~al.}(2019)\citenamefont {Island},
  \citenamefont {Cui}, \citenamefont {Lewandowski}, \citenamefont {Khoo},
  \citenamefont {Spanton}, \citenamefont {Zhou}, \citenamefont {Rhodes},
  \citenamefont {Hone}, \citenamefont {Taniguchi}, \citenamefont {Watanabe},
  \citenamefont {Levitov}, \citenamefont {Zaletel},\ and\ \citenamefont
  {Young}}]{island_spinorbit-driven_2019}%
  \BibitemOpen
  \bibfield  {author} {\bibinfo {author} {\bibfnamefont {J.~O.}\ \bibnamefont
  {Island}}, \bibinfo {author} {\bibfnamefont {X.}~\bibnamefont {Cui}},
  \bibinfo {author} {\bibfnamefont {C.}~\bibnamefont {Lewandowski}}, \bibinfo
  {author} {\bibfnamefont {J.~Y.}\ \bibnamefont {Khoo}}, \bibinfo {author}
  {\bibfnamefont {E.~M.}\ \bibnamefont {Spanton}}, \bibinfo {author}
  {\bibfnamefont {H.}~\bibnamefont {Zhou}}, \bibinfo {author} {\bibfnamefont
  {D.}~\bibnamefont {Rhodes}}, \bibinfo {author} {\bibfnamefont {J.~C.}\
  \bibnamefont {Hone}}, \bibinfo {author} {\bibfnamefont {T.}~\bibnamefont
  {Taniguchi}}, \bibinfo {author} {\bibfnamefont {K.}~\bibnamefont {Watanabe}},
  \bibinfo {author} {\bibfnamefont {L.~S.}\ \bibnamefont {Levitov}}, \bibinfo
  {author} {\bibfnamefont {M.~P.}\ \bibnamefont {Zaletel}}, \ and\ \bibinfo
  {author} {\bibfnamefont {A.~F.}\ \bibnamefont {Young}},\ }\href {\doibase
  10.1038/s41586-019-1304-2} {\bibfield  {journal} {\bibinfo  {journal}
  {Nature}\ ,\ \bibinfo {pages} {1}} (\bibinfo {year} {2019})}\BibitemShut
  {NoStop}%
\bibitem [{\citenamefont {Khoo}\ \emph {et~al.}(2017)\citenamefont {Khoo},
  \citenamefont {Morpurgo},\ and\ \citenamefont {Levitov}}]{khoo_-demand_2017}%
  \BibitemOpen
  \bibfield  {author} {\bibinfo {author} {\bibfnamefont {J.~Y.}\ \bibnamefont
  {Khoo}}, \bibinfo {author} {\bibfnamefont {A.~F.}\ \bibnamefont {Morpurgo}},
  \ and\ \bibinfo {author} {\bibfnamefont {L.}~\bibnamefont {Levitov}},\ }\href
  {\doibase 10.1021/acs.nanolett.7b03604} {\bibfield  {journal} {\bibinfo
  {journal} {Nano Lett.}\ }\textbf {\bibinfo {volume} {17}},\ \bibinfo {pages}
  {7003} (\bibinfo {year} {2017})}\BibitemShut {NoStop}%
\bibitem [{\citenamefont {Gmitra}\ and\ \citenamefont
  {Fabian}(2017)}]{gmitra_proximity_2017}%
  \BibitemOpen
  \bibfield  {author} {\bibinfo {author} {\bibfnamefont {M.}~\bibnamefont
  {Gmitra}}\ and\ \bibinfo {author} {\bibfnamefont {J.}~\bibnamefont
  {Fabian}},\ }\href {\doibase 10.1103/PhysRevLett.119.146401} {\bibfield
  {journal} {\bibinfo  {journal} {Phys. Rev. Lett.}\ }\textbf {\bibinfo
  {volume} {119}},\ \bibinfo {pages} {146401} (\bibinfo {year}
  {2017})}\BibitemShut {NoStop}%
\bibitem [{\citenamefont {Afzal}\ \emph {et~al.}(2018)\citenamefont {Afzal},
  \citenamefont {Khan}, \citenamefont {Nazir}, \citenamefont {Dastgeer},
  \citenamefont {Aftab}, \citenamefont {Akhtar}, \citenamefont {Seo},\ and\
  \citenamefont {Eom}}]{afzal_gate_2018}%
  \BibitemOpen
  \bibfield  {author} {\bibinfo {author} {\bibfnamefont {A.~M.}\ \bibnamefont
  {Afzal}}, \bibinfo {author} {\bibfnamefont {M.~F.}\ \bibnamefont {Khan}},
  \bibinfo {author} {\bibfnamefont {G.}~\bibnamefont {Nazir}}, \bibinfo
  {author} {\bibfnamefont {G.}~\bibnamefont {Dastgeer}}, \bibinfo {author}
  {\bibfnamefont {S.}~\bibnamefont {Aftab}}, \bibinfo {author} {\bibfnamefont
  {I.}~\bibnamefont {Akhtar}}, \bibinfo {author} {\bibfnamefont
  {Y.}~\bibnamefont {Seo}}, \ and\ \bibinfo {author} {\bibfnamefont
  {J.}~\bibnamefont {Eom}},\ }\href {\doibase 10.1038/s41598-018-21787-y}
  {\bibfield  {journal} {\bibinfo  {journal} {Sci. Rep.}\ }\textbf {\bibinfo
  {volume} {8}},\ \bibinfo {pages} {3412} (\bibinfo {year} {2018})}\BibitemShut
  {NoStop}%
\bibitem [{\citenamefont {Ye}\ \emph {et~al.}(2017)\citenamefont {Ye},
  \citenamefont {Yuan}, \citenamefont {Zhao},\ and\ \citenamefont
  {Guo}}]{ye_electric_2017}%
  \BibitemOpen
  \bibfield  {author} {\bibinfo {author} {\bibfnamefont {P.}~\bibnamefont
  {Ye}}, \bibinfo {author} {\bibfnamefont {R.~Y.}\ \bibnamefont {Yuan}},
  \bibinfo {author} {\bibfnamefont {X.}~\bibnamefont {Zhao}}, \ and\ \bibinfo
  {author} {\bibfnamefont {Y.}~\bibnamefont {Guo}},\ }\href {\doibase
  10.1063/1.4980109} {\bibfield  {journal} {\bibinfo  {journal} {J. Appl.
  Phys.}\ }\textbf {\bibinfo {volume} {121}},\ \bibinfo {pages} {144302}
  (\bibinfo {year} {2017})}\BibitemShut {NoStop}%
\bibitem [{\citenamefont {Omar}\ and\ \citenamefont {van
  Wees}(2017)}]{omar_graphene-$mathrmws_2$_2017}%
  \BibitemOpen
  \bibfield  {author} {\bibinfo {author} {\bibfnamefont {S.}~\bibnamefont
  {Omar}}\ and\ \bibinfo {author} {\bibfnamefont {B.~J.}\ \bibnamefont {van
  Wees}},\ }\href {\doibase 10.1103/PhysRevB.95.081404} {\bibfield  {journal}
  {\bibinfo  {journal} {Phys. Rev. B}\ }\textbf {\bibinfo {volume} {95}},\
  \bibinfo {pages} {081404(R)} (\bibinfo {year} {2017})}\BibitemShut {NoStop}%
\bibitem [{\citenamefont {Offidani}\ \emph {et~al.}(2017)\citenamefont
  {Offidani}, \citenamefont {Milletar\`{i}}, \citenamefont {Raimondi},\ and\
  \citenamefont {Ferreira}}]{offidani_optimal_2017}%
  \BibitemOpen
  \bibfield  {author} {\bibinfo {author} {\bibfnamefont {M.}~\bibnamefont
  {Offidani}}, \bibinfo {author} {\bibfnamefont {M.}~\bibnamefont
  {Milletar\`{i}}}, \bibinfo {author} {\bibfnamefont {R.}~\bibnamefont
  {Raimondi}}, \ and\ \bibinfo {author} {\bibfnamefont {A.}~\bibnamefont
  {Ferreira}},\ }\href {\doibase 10.1103/PhysRevLett.119.196801} {\bibfield
  {journal} {\bibinfo  {journal} {Phys. Rev. Lett.}\ }\textbf {\bibinfo
  {volume} {119}},\ \bibinfo {pages} {196801} (\bibinfo {year}
  {2017})}\BibitemShut {NoStop}%
\bibitem [{\citenamefont {Huang}\ \emph {et~al.}(2017)\citenamefont {Huang},
  \citenamefont {Chong},\ and\ \citenamefont
  {Cazalilla}}]{huang_anomalous_2017}%
  \BibitemOpen
  \bibfield  {author} {\bibinfo {author} {\bibfnamefont {C.}~\bibnamefont
  {Huang}}, \bibinfo {author} {\bibfnamefont {Y.D.}~\bibnamefont {Chong}}, \ and\
  \bibinfo {author} {\bibfnamefont {M.~A.}\ \bibnamefont {Cazalilla}},\ }\href
  {\doibase 10.1103/PhysRevLett.119.136804} {\bibfield  {journal} {\bibinfo
  {journal} {Phys. Rev. Lett.}\ }\textbf {\bibinfo {volume} {119}},\ \bibinfo
  {pages} {136804} (\bibinfo {year} {2017})}\BibitemShut {NoStop}%
\bibitem [{\citenamefont {Ando}\ and\ \citenamefont
  {Shiraishi}(2016)}]{ando_spin_2016}%
  \BibitemOpen
  \bibfield  {author} {\bibinfo {author} {\bibfnamefont {Y.}~\bibnamefont
  {Ando}}\ and\ \bibinfo {author} {\bibfnamefont {M.}~\bibnamefont
  {Shiraishi}},\ }\href {\doibase 10.7566/JPSJ.86.011001} {\bibfield  {journal}
  {\bibinfo  {journal} {J. Phys. Soc. Jpn.}\ }\textbf {\bibinfo {volume}
  {86}},\ \bibinfo {pages} {011001} (\bibinfo {year} {2016})}\BibitemShut
  {NoStop}%
\bibitem [{\citenamefont {Gurram}\ \emph {et~al.}(2018)\citenamefont {Gurram},
  \citenamefont {Omar},\ and\ \citenamefont {van
  Wees}}]{gurram_electrical_2018}%
  \BibitemOpen
  \bibfield  {author} {\bibinfo {author} {\bibfnamefont {M.}~\bibnamefont
  {Gurram}}, \bibinfo {author} {\bibfnamefont {S.}~\bibnamefont {Omar}}, \ and\
  \bibinfo {author} {\bibfnamefont {B.~J.}\ \bibnamefont {van Wees}},\ }\href
  {\doibase 10.1088/2053-1583/aac34d} {\bibfield  {journal} {\bibinfo
  {journal} {2D Mater.}\ }\textbf {\bibinfo {volume} {5}},\ \bibinfo {pages}
  {032004} (\bibinfo {year} {2018})}\BibitemShut {NoStop}%
\bibitem [{\citenamefont {Wang}\ \emph {et~al.}(2015)\citenamefont {Wang},
  \citenamefont {Ki}, \citenamefont {Chen}, \citenamefont {Berger},
  \citenamefont {MacDonald},\ and\ \citenamefont
  {Morpurgo}}]{wang_strong_2015}%
  \BibitemOpen
  \bibfield  {author} {\bibinfo {author} {\bibfnamefont {Z.}~\bibnamefont
  {Wang}}, \bibinfo {author} {\bibfnamefont {D.-K.}\ \bibnamefont {Ki}},
  \bibinfo {author} {\bibfnamefont {H.}~\bibnamefont {Chen}}, \bibinfo {author}
  {\bibfnamefont {H.}~\bibnamefont {Berger}}, \bibinfo {author} {\bibfnamefont
  {A.~H.}\ \bibnamefont {MacDonald}}, \ and\ \bibinfo {author} {\bibfnamefont
  {A.~F.}\ \bibnamefont {Morpurgo}},\ }\href {\doibase 10.1038/ncomms9339}
  {\bibfield  {journal} {\bibinfo  {journal} {Nat. Commun.}\ }\textbf {\bibinfo
  {volume} {6}},\ \bibinfo {pages} {9339} (\bibinfo {year} {2015})}\BibitemShut
  {NoStop}%
\bibitem [{\citenamefont {Ben\'{i}tez}\ \emph {et~al.}(2018)\citenamefont
  {Ben\'{i}tez}, \citenamefont {Sierra}, \citenamefont {Torres}, \citenamefont
  {Arrighi}, \citenamefont {Bonell}, \citenamefont {Costache},\ and\
  \citenamefont {Valenzuela}}]{benitez_strongly_2018}%
  \BibitemOpen
  \bibfield  {author} {\bibinfo {author} {\bibfnamefont {L.~A.}\ \bibnamefont
  {Ben\'{i}tez}}, \bibinfo {author} {\bibfnamefont {J.~F.}\ \bibnamefont
  {Sierra}}, \bibinfo {author} {\bibfnamefont {W.~S.}\ \bibnamefont {Torres}},
  \bibinfo {author} {\bibfnamefont {A.}~\bibnamefont {Arrighi}}, \bibinfo
  {author} {\bibfnamefont {F.}~\bibnamefont {Bonell}}, \bibinfo {author}
  {\bibfnamefont {M.~V.}\ \bibnamefont {Costache}}, \ and\ \bibinfo {author}
  {\bibfnamefont {S.~O.}\ \bibnamefont {Valenzuela}},\ }\href {\doibase
  10.1038/s41567-017-0019-2} {\bibfield  {journal} {\bibinfo  {journal} {Nat.
  Phys.}\ }\textbf {\bibinfo {volume} {14}},\ \bibinfo {pages} {303} (\bibinfo
  {year} {2018})}\BibitemShut {NoStop}%
\bibitem [{\citenamefont {Zhu}\ and\ \citenamefont
  {Kawakami}(2018)}]{zhu_modeling_2018}%
  \BibitemOpen
  \bibfield  {author} {\bibinfo {author} {\bibfnamefont {T.}~\bibnamefont
  {Zhu}}\ and\ \bibinfo {author} {\bibfnamefont {R.~K.}\ \bibnamefont
  {Kawakami}},\ }\href {\doibase 10.1103/PhysRevB.97.144413} {\bibfield
  {journal} {\bibinfo  {journal} {Phys. Rev. B}\ }\textbf {\bibinfo {volume}
  {97}},\ \bibinfo {pages} {144413} (\bibinfo {year} {2018})}\BibitemShut
  {NoStop}%
\bibitem [{\citenamefont {Tombros}\ \emph {et~al.}(2008)\citenamefont
  {Tombros}, \citenamefont {Tanabe}, \citenamefont {Veligura}, \citenamefont
  {J\'{o}zsa}, \citenamefont {Popinciuc}, \citenamefont {Jonkman},\ and\
  \citenamefont {van Wees}}]{tombros_anisotropic_2008}%
  \BibitemOpen
  \bibfield  {author} {\bibinfo {author} {\bibfnamefont {N.}~\bibnamefont
  {Tombros}}, \bibinfo {author} {\bibfnamefont {S.}~\bibnamefont {Tanabe}},
  \bibinfo {author} {\bibfnamefont {A.}~\bibnamefont {Veligura}}, \bibinfo
  {author} {\bibfnamefont {C.}~\bibnamefont {J\'{o}zsa}}, \bibinfo {author}
  {\bibfnamefont {M.}~\bibnamefont {Popinciuc}}, \bibinfo {author}
  {\bibfnamefont {H.~T.}\ \bibnamefont {Jonkman}}, \ and\ \bibinfo {author}
  {\bibfnamefont {B.~J.}\ \bibnamefont {van Wees}},\ }\href {\doibase
  10.1103/PhysRevLett.101.046601} {\bibfield  {journal} {\bibinfo  {journal}
  {Phys. Rev. Lett.}\ }\textbf {\bibinfo {volume} {101}},\ \bibinfo {pages}
  {046601} (\bibinfo {year} {2008})}\BibitemShut {NoStop}%
\bibitem [{\citenamefont {Popinciuc}\ \emph {et~al.}(2009)\citenamefont
  {Popinciuc}, \citenamefont {J\'{o}zsa}, \citenamefont {Zomer}, \citenamefont
  {Tombros}, \citenamefont {Veligura}, \citenamefont {Jonkman},\ and\
  \citenamefont {van Wees}}]{popinciuc_electronic_2009}%
  \BibitemOpen
  \bibfield  {author} {\bibinfo {author} {\bibfnamefont {M.}~\bibnamefont
  {Popinciuc}}, \bibinfo {author} {\bibfnamefont {C.}~\bibnamefont
  {J\'{o}zsa}}, \bibinfo {author} {\bibfnamefont {P.~J.}\ \bibnamefont
  {Zomer}}, \bibinfo {author} {\bibfnamefont {N.}~\bibnamefont {Tombros}},
  \bibinfo {author} {\bibfnamefont {A.}~\bibnamefont {Veligura}}, \bibinfo
  {author} {\bibfnamefont {H.~T.}\ \bibnamefont {Jonkman}}, \ and\ \bibinfo
  {author} {\bibfnamefont {B.~J.}\ \bibnamefont {van Wees}},\ }\href {\doibase
  10.1103/PhysRevB.80.214427} {\bibfield  {journal} {\bibinfo  {journal} {Phys.
  Rev. B}\ }\textbf {\bibinfo {volume} {80}},\ \bibinfo {pages} {214427}
  (\bibinfo {year} {2009})}\BibitemShut {NoStop}%
\bibitem [{\citenamefont {Guimar\~{a}es}\ \emph {et~al.}(2014)\citenamefont
  {Guimar\~{a}es}, \citenamefont {Zomer}, \citenamefont {Ingla-Ayn\'{e}s},
  \citenamefont {Brant}, \citenamefont {Tombros},\ and\ \citenamefont {van
  Wees}}]{guimaraes_controlling_2014}%
  \BibitemOpen
  \bibfield  {author} {\bibinfo {author} {\bibfnamefont {M.H.D.}~\bibnamefont
  {Guimar\~{a}es}}, \bibinfo {author} {\bibfnamefont {P.J.}~\bibnamefont
  {Zomer}}, \bibinfo {author} {\bibfnamefont {J.}~\bibnamefont
  {Ingla-Ayn\'{e}s}}, \bibinfo {author} {\bibfnamefont {J.C.}~\bibnamefont
  {Brant}}, \bibinfo {author} {\bibfnamefont {N.}~\bibnamefont {Tombros}}, \
  and\ \bibinfo {author} {\bibfnamefont {B.J.}~\bibnamefont {van Wees}},\ }\href
  {\doibase 10.1103/PhysRevLett.113.086602} {\bibfield  {journal} {\bibinfo
  {journal} {Phys. Rev. Lett.}\ }\textbf {\bibinfo {volume} {113}},\ \bibinfo
  {pages} {086602} (\bibinfo {year} {2014})}\BibitemShut {NoStop}%
\bibitem [{\citenamefont {Avsar}\ \emph {et~al.}(2017)\citenamefont {Avsar},
  \citenamefont {Unuchek}, \citenamefont {Liu}, \citenamefont {Sanchez},
  \citenamefont {Watanabe}, \citenamefont {Taniguchi}, \citenamefont
  {\"{O}zyilmaz},\ and\ \citenamefont {Kis}}]{avsar_optospintronics_2017}%
  \BibitemOpen
  \bibfield  {author} {\bibinfo {author} {\bibfnamefont {A.}~\bibnamefont
  {Avsar}}, \bibinfo {author} {\bibfnamefont {D.}~\bibnamefont {Unuchek}},
  \bibinfo {author} {\bibfnamefont {J.}~\bibnamefont {Liu}}, \bibinfo {author}
  {\bibfnamefont {O.~L.}\ \bibnamefont {Sanchez}}, \bibinfo {author}
  {\bibfnamefont {K.}~\bibnamefont {Watanabe}}, \bibinfo {author}
  {\bibfnamefont {T.}~\bibnamefont {Taniguchi}}, \bibinfo {author}
  {\bibfnamefont {B.}~\bibnamefont {\"{O}zyilmaz}}, \ and\ \bibinfo {author}
  {\bibfnamefont {A.}~\bibnamefont {Kis}},\ }\href {\doibase
  10.1021/acsnano.7b06800} {\bibfield  {journal} {\bibinfo  {journal} {ACS
  Nano}\ }\textbf {\bibinfo {volume} {11}},\ \bibinfo {pages} {11678} (\bibinfo
  {year} {2017})}\BibitemShut {NoStop}%
\bibitem [{\citenamefont {Luo}\ \emph {et~al.}(2017)\citenamefont {Luo},
  \citenamefont {Xu}, \citenamefont {Zhu}, \citenamefont {Wu}, \citenamefont
  {McCormick}, \citenamefont {Zhan}, \citenamefont {Neupane},\ and\
  \citenamefont {Kawakami}}]{luo_opto-valleytronic_2017}%
  \BibitemOpen
  \bibfield  {author} {\bibinfo {author} {\bibfnamefont {Y.~K.}\ \bibnamefont
  {Luo}}, \bibinfo {author} {\bibfnamefont {J.}~\bibnamefont {Xu}}, \bibinfo
  {author} {\bibfnamefont {T.}~\bibnamefont {Zhu}}, \bibinfo {author}
  {\bibfnamefont {G.}~\bibnamefont {Wu}}, \bibinfo {author} {\bibfnamefont
  {E.~J.}\ \bibnamefont {McCormick}}, \bibinfo {author} {\bibfnamefont
  {W.}~\bibnamefont {Zhan}}, \bibinfo {author} {\bibfnamefont {M.~R.}\
  \bibnamefont {Neupane}}, \ and\ \bibinfo {author} {\bibfnamefont {R.~K.}\
  \bibnamefont {Kawakami}},\ }\href {\doibase 10.1021/acs.nanolett.7b01393}
  {\bibfield  {journal} {\bibinfo  {journal} {Nano Lett.}\ }\textbf {\bibinfo
  {volume} {17}},\ \bibinfo {pages} {3877} (\bibinfo {year}
  {2017})}\BibitemShut {NoStop}%
\bibitem [{\citenamefont {Zomer}\ \emph {et~al.}(2014)\citenamefont {Zomer},
  \citenamefont {Guimar\~{a}es}, \citenamefont {Brant}, \citenamefont
  {Tombros},\ and\ \citenamefont {van Wees}}]{zomer_fast_2014}%
  \BibitemOpen
  \bibfield  {author} {\bibinfo {author} {\bibfnamefont {P.~J.}\ \bibnamefont
  {Zomer}}, \bibinfo {author} {\bibfnamefont {M.~H.~D.}\ \bibnamefont
  {Guimar\~{a}es}}, \bibinfo {author} {\bibfnamefont {J.~C.}\ \bibnamefont
  {Brant}}, \bibinfo {author} {\bibfnamefont {N.}~\bibnamefont {Tombros}}, \
  and\ \bibinfo {author} {\bibfnamefont {B.~J.}\ \bibnamefont {van Wees}},\
  }\href {\doibase 10.1063/1.4886096} {\bibfield  {journal} {\bibinfo
  {journal} {Appl. Phys. Lett.}\ }\textbf {\bibinfo {volume} {105}},\ \bibinfo
  {pages} {013101} (\bibinfo {year} {2014})}\BibitemShut {NoStop}%
\bibitem [{\citenamefont {Yan}\ \emph {et~al.}(2016)\citenamefont {Yan},
  \citenamefont {Txoperena}, \citenamefont {Llopis}, \citenamefont {Dery},
  \citenamefont {Hueso},\ and\ \citenamefont
  {Casanova}}]{yan_two-dimensional_2016}%
  \BibitemOpen
  \bibfield  {author} {\bibinfo {author} {\bibfnamefont {W.}~\bibnamefont
  {Yan}}, \bibinfo {author} {\bibfnamefont {O.}~\bibnamefont {Txoperena}},
  \bibinfo {author} {\bibfnamefont {R.}~\bibnamefont {Llopis}}, \bibinfo
  {author} {\bibfnamefont {H.}~\bibnamefont {Dery}}, \bibinfo {author}
  {\bibfnamefont {L.~E.}\ \bibnamefont {Hueso}}, \ and\ \bibinfo {author}
  {\bibfnamefont {F.}~\bibnamefont {Casanova}},\ }\href {\doibase
  10.1038/ncomms13372} {\bibfield  {journal} {\bibinfo  {journal} {Nat.
  Commun.}\ }\textbf {\bibinfo {volume} {7}},\ \bibinfo {pages} {13372}
  (\bibinfo {year} {2016})}\BibitemShut {NoStop}%
\bibitem [{\citenamefont {Dankert}\ and\ \citenamefont
  {Dash}(2017)}]{dankert_electrical_2017}%
  \BibitemOpen
  \bibfield  {author} {\bibinfo {author} {\bibfnamefont {A.}~\bibnamefont
  {Dankert}}\ and\ \bibinfo {author} {\bibfnamefont {S.~P.}\ \bibnamefont
  {Dash}},\ }\href {\doibase 10.1038/ncomms16093} {\bibfield  {journal}
  {\bibinfo  {journal} {Nat. Commun.}\ }\textbf {\bibinfo {volume} {8}},\
  \bibinfo {pages} {16093} (\bibinfo {year} {2017})}\BibitemShut {NoStop}%
\bibitem [{\citenamefont {Gurram}\ \emph {et~al.}(2017)\citenamefont {Gurram},
  \citenamefont {Omar},\ and\ \citenamefont {van Wees}}]{gurram_bias_2017}%
  \BibitemOpen
  \bibfield  {author} {\bibinfo {author} {\bibfnamefont {M.}~\bibnamefont
  {Gurram}}, \bibinfo {author} {\bibfnamefont {S.}~\bibnamefont {Omar}}, \ and\
  \bibinfo {author} {\bibfnamefont {B.~J.}\ \bibnamefont {van Wees}},\ }\href
  {\doibase 10.1038/s41467-017-00317-w} {\bibfield  {journal} {\bibinfo
  {journal} {Nat. Commun.}\ }\textbf {\bibinfo {volume} {8}},\ \bibinfo {pages}
  {248} (\bibinfo {year} {2017})}\BibitemShut {NoStop}%
\bibitem [{\citenamefont {Ingla-Ayn\'{e}s}\ \emph {et~al.}(2015)\citenamefont
  {Ingla-Ayn\'{e}s}, \citenamefont {Guimar\~{a}es}, \citenamefont {Meijerink},
  \citenamefont {Zomer},\ and\ \citenamefont {van
  Wees}}]{ingla-aynes_$24ensuremath-ensuremathmumathrmm$_2015}%
  \BibitemOpen
  \bibfield  {author} {\bibinfo {author} {\bibfnamefont {J.}~\bibnamefont
  {Ingla-Ayn\'{e}s}}, \bibinfo {author} {\bibfnamefont {M.~H.~D.}\ \bibnamefont
  {Guimar\~{a}es}}, \bibinfo {author} {\bibfnamefont {R.~J.}\ \bibnamefont
  {Meijerink}}, \bibinfo {author} {\bibfnamefont {P.~J.}\ \bibnamefont
  {Zomer}}, \ and\ \bibinfo {author} {\bibfnamefont {B.~J.}\ \bibnamefont {van
  Wees}},\ }\href {\doibase 10.1103/PhysRevB.92.201410} {\bibfield  {journal}
  {\bibinfo  {journal} {Phys. Rev. B}\ }\textbf {\bibinfo {volume} {92}},\
  \bibinfo {pages} {201410(R)} (\bibinfo {year} {2015})}\BibitemShut {NoStop}%
\bibitem [{\citenamefont {Dr\"{o}geler}\ \emph {et~al.}(2016)\citenamefont
  {Dr\"{o}geler}, \citenamefont {Franzen}, \citenamefont {Volmer},
  \citenamefont {Pohlmann}, \citenamefont {Banszerus}, \citenamefont {Wolter},
  \citenamefont {Watanabe}, \citenamefont {Taniguchi}, \citenamefont
  {Stampfer},\ and\ \citenamefont {Beschoten}}]{drogeler_spin_2016}%
  \BibitemOpen
  \bibfield  {author} {\bibinfo {author} {\bibfnamefont {M.}~\bibnamefont
  {Dr\"{o}geler}}, \bibinfo {author} {\bibfnamefont {C.}~\bibnamefont
  {Franzen}}, \bibinfo {author} {\bibfnamefont {F.}~\bibnamefont {Volmer}},
  \bibinfo {author} {\bibfnamefont {T.}~\bibnamefont {Pohlmann}}, \bibinfo
  {author} {\bibfnamefont {L.}~\bibnamefont {Banszerus}}, \bibinfo {author}
  {\bibfnamefont {M.}~\bibnamefont {Wolter}}, \bibinfo {author} {\bibfnamefont
  {K.}~\bibnamefont {Watanabe}}, \bibinfo {author} {\bibfnamefont
  {T.}~\bibnamefont {Taniguchi}}, \bibinfo {author} {\bibfnamefont
  {C.}~\bibnamefont {Stampfer}}, \ and\ \bibinfo {author} {\bibfnamefont
  {B.}~\bibnamefont {Beschoten}},\ }\href {\doibase
  10.1021/acs.nanolett.6b00497} {\bibfield  {journal} {\bibinfo  {journal}
  {Nano Lett.}\ }\textbf {\bibinfo {volume} {16}},\ \bibinfo {pages} {3533}
  (\bibinfo {year} {2016})}\BibitemShut {NoStop}%
\end{thebibliography}

\begin{thebibliography}{9}%
\makeatletter
\providecommand \@ifxundefined [1]{%
 \@ifx{#1\undefined}
}%
\providecommand \@ifnum [1]{%
 \ifnum #1\expandafter \@firstoftwo
 \else \expandafter \@secondoftwo
 \fi
}%
\providecommand \@ifx [1]{%
 \ifx #1\expandafter \@firstoftwo
 \else \expandafter \@secondoftwo
 \fi
}%
\providecommand \natexlab [1]{#1}%
\providecommand \enquote  [1]{``#1''}%
\providecommand \bibnamefont  [1]{#1}%
\providecommand \bibfnamefont [1]{#1}%
\providecommand \citenamefont [1]{#1}%
\providecommand \href@noop [0]{\@secondoftwo}%
\providecommand \href [0]{\begingroup \@sanitize@url \@href}%
\providecommand \@href[1]{\@@startlink{#1}\@@href}%
\providecommand \@@href[1]{\endgroup#1\@@endlink}%
\providecommand \@sanitize@url [0]{\catcode `\\12\catcode `\$12\catcode
  `\&12\catcode `\#12\catcode `\^12\catcode `\_12\catcode `\%12\relax}%
\providecommand \@@startlink[1]{}%
\providecommand \@@endlink[0]{}%
\providecommand \url  [0]{\begingroup\@sanitize@url \@url }%
\providecommand \@url [1]{\endgroup\@href {#1}{\urlprefix }}%
\providecommand \urlprefix  [0]{URL }%
\providecommand \Eprint [0]{\href }%
\providecommand \doibase [0]{http://dx.doi.org/}%
\providecommand \selectlanguage [0]{\@gobble}%
\providecommand \bibinfo  [0]{\@secondoftwo}%
\providecommand \bibfield  [0]{\@secondoftwo}%
\providecommand \translation [1]{[#1]}%
\providecommand \BibitemOpen [0]{}%
\providecommand \bibitemStop [0]{}%
\providecommand \bibitemNoStop [0]{.\EOS\space}%
\providecommand \EOS [0]{\spacefactor3000\relax}%
\providecommand \BibitemShut  [1]{\csname bibitem#1\endcsname}%
\let\auto@bib@innerbib\@empty
%</preamble>
\bibitem [{\citenamefont {Gurram}\ \emph {et~al.}(2016)\citenamefont {Gurram},
  \citenamefont {Omar}, \citenamefont {Zihlmann}, \citenamefont {Makk},
  \citenamefont {Sch\"{o}nenberger},\ and\ \citenamefont {van
  Wees}}]{gurram_spin_2016}%
  \BibitemOpen
  \bibfield  {author} {\bibinfo {author} {\bibfnamefont {M.}~\bibnamefont
  {Gurram}}, \bibinfo {author} {\bibfnamefont {S.}~\bibnamefont {Omar}},
  \bibinfo {author} {\bibfnamefont {S.}~\bibnamefont {Zihlmann}}, \bibinfo
  {author} {\bibfnamefont {P.}~\bibnamefont {Makk}}, \bibinfo {author}
  {\bibfnamefont {C.}~\bibnamefont {Sch\"{o}nenberger}}, \ and\ \bibinfo
  {author} {\bibfnamefont {B.~J.}\ \bibnamefont {van Wees}},\ }\href {\doibase
  10.1103/PhysRevB.93.115441} {\bibfield  {journal} {\bibinfo  {journal} {Phys.
  Rev. B}\ }\textbf {\bibinfo {volume} {93}},\ \bibinfo {pages} {115441}
  (\bibinfo {year} {2016})}\BibitemShut {NoStop}%
\bibitem [{\citenamefont {Zomer}\ \emph {et~al.}(2014)\citenamefont {Zomer},
  \citenamefont {Guimar\~{a}es}, \citenamefont {Brant}, \citenamefont
  {Tombros},\ and\ \citenamefont {van Wees}}]{zomer_fast_2014}%
  \BibitemOpen
  \bibfield  {author} {\bibinfo {author} {\bibfnamefont {P.~J.}\ \bibnamefont
  {Zomer}}, \bibinfo {author} {\bibfnamefont {M.~H.~D.}\ \bibnamefont
  {Guimar\~{a}es}}, \bibinfo {author} {\bibfnamefont {J.~C.}\ \bibnamefont
  {Brant}}, \bibinfo {author} {\bibfnamefont {N.}~\bibnamefont {Tombros}}, \
  and\ \bibinfo {author} {\bibfnamefont {B.~J.}\ \bibnamefont {van Wees}},\
  }\href {\doibase 10.1063/1.4886096} {\bibfield  {journal} {\bibinfo
  {journal} {Appl. Phys. Lett.}\ }\textbf {\bibinfo {volume} {105}},\ \bibinfo
  {pages} {013101} (\bibinfo {year} {2014})}\BibitemShut {NoStop}%
\bibitem [{\citenamefont {Maassen}\ \emph {et~al.}(2012)\citenamefont
  {Maassen}, \citenamefont {Vera-Marun}, \citenamefont {Guimar\~{a}es},\ and\
  \citenamefont {van Wees}}]{maassen_contact-induced_2012}%
  \BibitemOpen
  \bibfield  {author} {\bibinfo {author} {\bibfnamefont {T.}~\bibnamefont
  {Maassen}}, \bibinfo {author} {\bibfnamefont {I.~J.}\ \bibnamefont
  {Vera-Marun}}, \bibinfo {author} {\bibfnamefont {M.~H.~D.}\ \bibnamefont
  {Guimar\~{a}es}}, \ and\ \bibinfo {author} {\bibfnamefont {B.~J.}\
  \bibnamefont {van Wees}},\ }\href {\doibase 10.1103/PhysRevB.86.235408}
  {\bibfield  {journal} {\bibinfo  {journal} {Phys. Rev. B}\ }\textbf {\bibinfo
  {volume} {86}},\ \bibinfo {pages} {235408} (\bibinfo {year}
  {2012})}\BibitemShut {NoStop}%
\bibitem [{\citenamefont {Omar}\ and\ \citenamefont {van
  Wees}(2017)}]{omar_graphene-$mathrmws_2$_2017}%
  \BibitemOpen
  \bibfield  {author} {\bibinfo {author} {\bibfnamefont {S.}~\bibnamefont
  {Omar}}\ and\ \bibinfo {author} {\bibfnamefont {B.~J.}\ \bibnamefont {van
  Wees}},\ }\href {\doibase 10.1103/PhysRevB.95.081404} {\bibfield  {journal}
  {\bibinfo  {journal} {Phys. Rev. B}\ }\textbf {\bibinfo {volume} {95}},\
  \bibinfo {pages} {081404} (\bibinfo {year} {2017})}\BibitemShut {NoStop}%
\bibitem [{\citenamefont {Kittel}(2004)}]{kittel_introduction_2004}%
  \BibitemOpen
  \bibfield  {author} {\bibinfo {author} {\bibfnamefont {C.}~\bibnamefont
  {Kittel}},\ }\href
  {https://www.wiley.com/en-us/Introduction+to+Solid+State+Physics%2C+8th+Edition-p-9780471415268}
  {\emph {\bibinfo {title} {Introduction to {Solid} {State} {Physics}}}}\
  (\bibinfo  {publisher} {Wiley},\ \bibinfo {year} {2004})\BibitemShut
  {NoStop}%
\bibitem [{\citenamefont {Raes}\ \emph {et~al.}(2016)\citenamefont {Raes},
  \citenamefont {Scheerder}, \citenamefont {Costache}, \citenamefont {Bonell},
  \citenamefont {Sierra}, \citenamefont {Cuppens}, \citenamefont {Van~de
  Vondel},\ and\ \citenamefont {Valenzuela}}]{raes_determination_2016}%
  \BibitemOpen
  \bibfield  {author} {\bibinfo {author} {\bibfnamefont {B.}~\bibnamefont
  {Raes}}, \bibinfo {author} {\bibfnamefont {J.~E.}\ \bibnamefont {Scheerder}},
  \bibinfo {author} {\bibfnamefont {M.~V.}\ \bibnamefont {Costache}}, \bibinfo
  {author} {\bibfnamefont {F.}~\bibnamefont {Bonell}}, \bibinfo {author}
  {\bibfnamefont {J.~F.}\ \bibnamefont {Sierra}}, \bibinfo {author}
  {\bibfnamefont {J.}~\bibnamefont {Cuppens}}, \bibinfo {author} {\bibfnamefont
  {J.}~\bibnamefont {Van~de Vondel}}, \ and\ \bibinfo {author} {\bibfnamefont
  {S.~O.}\ \bibnamefont {Valenzuela}},\ }\href {\doibase 10.1038/ncomms11444}
  {\bibfield  {journal} {\bibinfo  {journal} {Nat. Commun.}\ }\textbf {\bibinfo
  {volume} {7}},\ \bibinfo {pages} {11444} (\bibinfo {year}
  {2016})}\BibitemShut {NoStop}%
\bibitem [{\citenamefont {Cummings}\ \emph {et~al.}(2017)\citenamefont
  {Cummings}, \citenamefont {Garcia}, \citenamefont {Fabian},\ and\
  \citenamefont {Roche}}]{cummings_giant_2017}%
  \BibitemOpen
  \bibfield  {author} {\bibinfo {author} {\bibfnamefont {A.~W.}\ \bibnamefont
  {Cummings}}, \bibinfo {author} {\bibfnamefont {J.~H.}\ \bibnamefont
  {Garcia}}, \bibinfo {author} {\bibfnamefont {J.}~\bibnamefont {Fabian}}, \
  and\ \bibinfo {author} {\bibfnamefont {S.}~\bibnamefont {Roche}},\ }\href
  {\doibase 10.1103/PhysRevLett.119.206601} {\bibfield  {journal} {\bibinfo
  {journal} {Phys. Rev. Lett.}\ }\textbf {\bibinfo {volume} {119}},\ \bibinfo
  {pages} {206601} (\bibinfo {year} {2017})}\BibitemShut {NoStop}%
\bibitem [{\citenamefont {Omar}\ and\ \citenamefont {van
  Wees}(2018)}]{omar_spin_2018}%
  \BibitemOpen
  \bibfield  {author} {\bibinfo {author} {\bibfnamefont {S.}~\bibnamefont
  {Omar}}\ and\ \bibinfo {author} {\bibfnamefont {B.~J.}\ \bibnamefont {van
  Wees}},\ }\href {\doibase 10.1103/PhysRevB.97.045414} {\bibfield  {journal}
  {\bibinfo  {journal} {Phys. Rev. B}\ }\textbf {\bibinfo {volume} {97}},\
  \bibinfo {pages} {045414} (\bibinfo {year} {2018})}\BibitemShut {NoStop}%
\bibitem [{\citenamefont {Zihlmann}\ \emph {et~al.}(2018)\citenamefont
  {Zihlmann}, \citenamefont {Cummings}, \citenamefont {Garcia}, \citenamefont
  {Kedves}, \citenamefont {Watanabe}, \citenamefont {Taniguchi}, \citenamefont
  {Sch\"{o}nenberger},\ and\ \citenamefont {Makk}}]{zihlmann_large_2018}%
  \BibitemOpen
  \bibfield  {author} {\bibinfo {author} {\bibfnamefont {S.}~\bibnamefont
  {Zihlmann}}, \bibinfo {author} {\bibfnamefont {A.~W.}\ \bibnamefont
  {Cummings}}, \bibinfo {author} {\bibfnamefont {J.~H.}\ \bibnamefont
  {Garcia}}, \bibinfo {author} {\bibfnamefont {M.}~\bibnamefont {Kedves}},
  \bibinfo {author} {\bibfnamefont {K.}~\bibnamefont {Watanabe}}, \bibinfo
  {author} {\bibfnamefont {T.}~\bibnamefont {Taniguchi}}, \bibinfo {author}
  {\bibfnamefont {C.}~\bibnamefont {Sch\"{o}nenberger}}, \ and\ \bibinfo
  {author} {\bibfnamefont {P.}~\bibnamefont {Makk}},\ }\href {\doibase
  10.1103/PhysRevB.97.075434} {\bibfield  {journal} {\bibinfo  {journal} {Phys.
  Rev. B}\ }\textbf {\bibinfo {volume} {97}},\ \bibinfo {pages} {075434}
  (\bibinfo {year} {2018})}\BibitemShut {NoStop}%
\end{thebibliography}

%

\end{document}